\documentclass[11pt,a4paper]{article}
\pdfoutput=1
\usepackage{jheppub}

\usepackage{multirow, graphicx,amssymb,url,mathrsfs,amsmath}
\usepackage{wrapfig,boxedminipage,setspace,subfigure,epsfig}
\usepackage{amsxtra,amstext,latexsym,dsfont,amsfonts}
\usepackage{color}
\usepackage[dvipsnames]{xcolor}
\usepackage{float}
\usepackage{slashed,comment}
\usepackage{contour}

\usepackage{tikz}
\usetikzlibrary{calc,patterns,angles,quotes}
\usetikzlibrary{decorations.pathreplacing,decorations.markings,snakes}

\usepackage{tabularx, array}
\newcolumntype{L}[1]{>{\raggedright\arraybackslash}p{#1}}
\newcolumntype{C}[1]{>{\centering\arraybackslash}p{#1}}
\newcolumntype{R}[1]{>{\raggedleft\arraybackslash}p{#1}}

\newcommand{\dd}{\mathrm{d}}

\usepackage{xparse}

\title{AdS/Deep-Learning made easy II: neural network-based approaches to holography and inverse problems}

\author[a,b]{Hyun-Sik Jeong,}
\author[c]{Hanse Kim,}
\author[d,e]{Keun-Young Kim,}
\author[d]{Gaya Yun,}
\author[d]{Hyeonwoo Yu,}
\author[d]{and Kwan Yun}

\emailAdd{hyunsik.jeong@apctp.org}
\emailAdd{hkim2389@wisc.edu}
\emailAdd{fortoe@gist.ac.kr}
\emailAdd{gayayun121@gm.gist.ac.kr}
\emailAdd{hyeonuyu759@gm.gist.ac.kr}
\emailAdd{ludibriphy70@gm.gist.ac.kr}

\preprint{\texttt{APCTP Pre2025 - 024}}

\affiliation[a]{Asia Pacific Center for Theoretical Physics, Pohang 37673, Korea}
\affiliation[b]{Department of Physics, Pohang University of Science and Technology, Pohang 37673, Korea}
\affiliation[c]{Department of Physics, University of Wisconsin-Madison, \\ 1150 University Avenue, Madison WI, United States}
\affiliation[d]{Department of Physics and Photon Science, Gwangju Institute of Science and Technology, \\
123 Cheomdan-gwagiro, Gwangju 61005, Korea}
\affiliation[e]{Research Center for Photon Science Technology, Gwangju Institute of Science and Technology, \\
123 Cheomdan-gwagiro, Gwangju 61005, Korea}

\abstract{We apply physics-informed machine learning (PIML) to solve inverse problems in holography and classical mechanics, focusing on neural ordinary differential equations (Neural ODEs) and physics-informed neural networks (PINNs) for solving non-linear differential equations of motion. First, we introduce holographic inverse problems and demonstrate how PIML can reconstruct bulk spacetime and effective potentials from boundary quantum data. To illustrate this, two case studies are explored: the QCD equation of state in holographic QCD and $T$-linear resistivity in holographic strange metals. Additionally, we explicitly show how such holographic problems can be analogized to inverse problems in classical mechanics, modeling frictional forces with neural networks. We also explore Kolmogorov-Arnold Networks (KANs) as an alternative to traditional neural networks, offering more efficient solutions in certain cases. This manuscript aim to provide a systematic framework for using neural networks in inverse problems, serving as a comprehensive reference for researchers in machine learning for high-energy physics, with methodologies that also have broader applications in mathematics, engineering, and the natural sciences.}

\begin{document}
\maketitle

%
\section{Introduction}\label{}
Scientific Machine Learning (SciML) has emerged as a rapidly expanding discipline with broad applications across scientific domains~\cite{Baker2019,Willard:2020aa,Cuomo:2022aa}. The central aim of SciML is to integrate established scientific principles with modern machine learning techniques, generating models that are not only data-driven but also informed by prior physical or theoretical knowledge. Within this context, physics-informed machine learning (PIML)~\cite{Karniadakis_2021,Sharma_2023,Toscano:2024aa,Ahmadi:2025aa} has developed as a prominent subfield that explicitly incorporates physical laws into the learning process.

Neural networks (NNs)~\cite{LeCun_2015,SCHMIDHUBER201585} have been particularly instrumental in advancing PIML due to their capacity to approximate complex non-linear relationships. They have been successfully applied to a wide range of physics problems, including nuclear physics~\cite{Boehnlein:2021eym}, quantum many-body systems~\cite{Carleo:2017aa,Bedolla-Montiel:2020rio}, and even string theory~\cite{Ruehle:2020jrk}. For comprehensive reviews of machine learning applications in physics, we refer the readers to \cite{Carleo:2019ptp,Tanaka_2021,Halverson:2024aa}. In what follows, we briefly outline the basic neural network architecture used in this manuscript and its connection to physics-informed frameworks.

\begin{figure}
    \centering
    \includegraphics[width=0.7\linewidth]{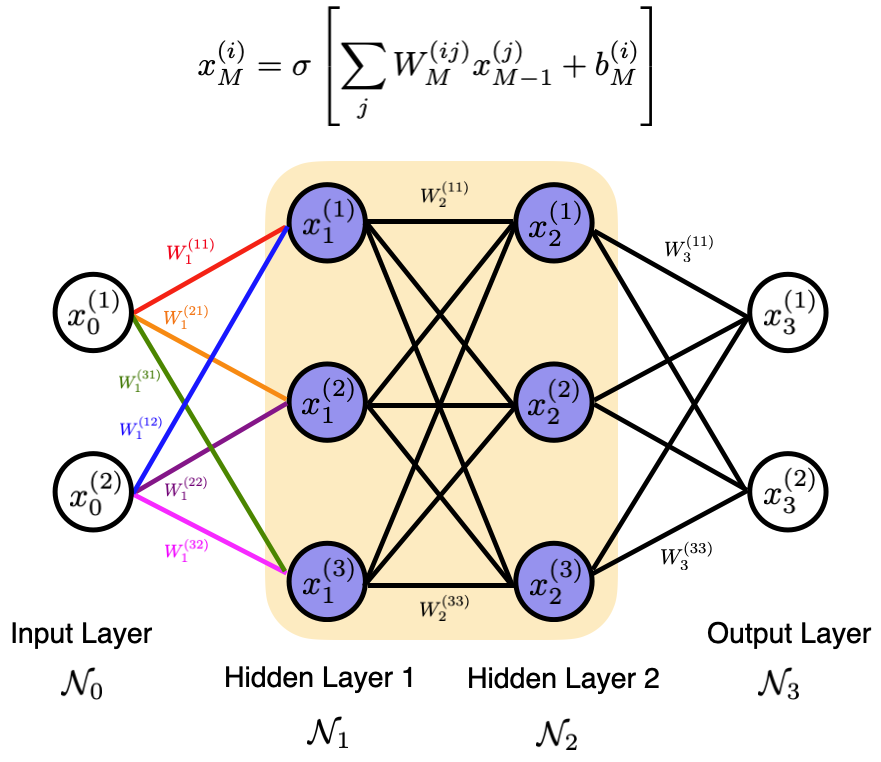}
    \caption{A schematic architecture of a deep neural network with two hidden layers. Each weight $W^{(ij)}_m$ connects neuron $x^{(i)}_{m-1}$ in layer $\mathcal{N}_{m-1}$ to neuron $x^{(i)}_m$ in $\mathcal{N}_m$. Here, $x^{(a)}_b$ means the $a$-th neuron in the $b$-th layer.}\label{ARCH}
\end{figure}

\paragraph{Neural network architecture.}
A neural network is a composition of functions organized in layers. Let the input layer be $\mathcal{N}_0(x)=x$ and the output layer $\mathcal{N}_M(x)$, where the total number of layers is $(M+1)$. Each layer $\mathcal{N}_m$ is parameterized by a weight matrix $W_m$ and a bias vector $b_m$. The hidden layers transform the input recursively as
\begin{equation}
\mathcal{N}_{m}(x)=\sigma \left[ W_{m} \, \mathcal{N}_{m-1} (x)+b_m \right]  \,, \quad (1\leq m < M) 
\end{equation}
where $\sigma$ denotes an activation function such as $\tanh(x)$, ReLU, LeakyReLU or SoftPlus. The final output layer $\mathcal{N}_{M}(x)$ is given as
\begin{equation}\label{DNN2}
    \mathcal{N}_{M}(x; \theta) = W_{M}  \, \mathcal{N}_{M-1}(x) + b_M=(\mathcal{N}_M \circ \mathcal{N}_{M-1} \cdots \mathcal{N}_{0})(x) \,,
\end{equation}
where the last equality is the composition of non-linear functions, $\circ$ denotes the function composition, and $\theta=\{W_{m}, b_m \}$ represents the collective set of learning parameters. Figure \ref{ARCH} illustrates this architecture schematically.

Each neuron processes a weighted sum of its inputs, applies a non-linear transformation, and transmits the result to the next layer. By adjusting weights and biases through optimization, NNs learn to approximate mappings between inputs and outputs, effectively modeling highly non-linear systems. The universal approximation theorem~\cite{HORNIK1989359} formalizes this capability, guaranteeing that for any continuous function $f(x)$ and any $\epsilon > 0$, there exists a network $\mathcal{N}_M(x; \theta)$ satisfying $|f(x) - \mathcal{N}_M(x; \theta)| < \epsilon$. This property underpins the ability of neural networks to serve as flexible function approximators, particularly useful for solving complex physical problems where analytical or numerical solutions are intractable.

\paragraph{PIML with neural networks.}
Traditional numerical solvers for differential equations become unstable, slow, or inaccurate when equations are high-dimensional (systems with many variables or equations), stiff (some parts of equation evolve much faster than others), or chaotic systems (highly sensitive to initial conditions). Neural networks overcome these limitations by learning implicit functional representations that approximate the solution of (ordinary or partial) differential equations. PIML frameworks harness this expressive capacity while embedding known physical constraints, such as conservation laws or boundary conditions, directly into the training objective.

In general, the loss function $L(\theta)$ is constructed to penalize violations of both empirical data and physical laws. The network parameters are optimized iteratively as
\begin{equation}
\theta^{j+1} = \theta^j - l_r \nabla_\theta L(\theta^j) \,,
\end{equation}
where $l_r$ is the learning rate and $j$ indexes the training iteration (epoch). Automatic differentiation, implemented efficiently in libraries such as \texttt{PyTorch} and \texttt{TensorFlow}, enables the computation of $\nabla_\theta L$ with respect to all parameters, making PIML methods computationally practical and feasible.

Several optimization algorithms can further be employed to minimize the loss function. In our work, we first use the Adam optimizer~\cite{Kingma:2014vow} to efficiently approach the vicinity of a minimum, and subsequently apply the L-BFGS algorithm~\cite{Liu1989} to refine the solution and achieve higher precision.

Two major methodological paradigms dominate the field:
(I) Neural Ordinary Differential Equations (Neural ODEs)~\cite{Chen:2018wjc,Dupont:2019aa,Massaroli:2020aa,Yan:2019aa}:
Neural networks parameterize the dynamics of differential equations, and gradients are propagated through the differential system itself to optimize parameters. (II) Physics-Informed Neural Networks (PINNs)~\cite{RAISSI2019686,HBWJ2023,Raissi:2017aa,Raissi:2017ab}: PINNs replace the unknown function in differential equations with a neural network. The loss function enforces the governing equation and boundary/initial conditions, ensuring that minimizing the loss yields an approximate physical solution.\footnote{Related approaches include neural operators~\cite{Lu_2021}, which learn mappings between function spaces, and universal differential equations~\cite{Rackauckas:2020aa}, where only certain selected terms in equations are replaced by neural components.} 

In practice, within Neural ODEs, the differential equation is solved using a conventional ODE solver, but the unknown target function is parameterized by a neural network. In contrast, PINNs incorporate the differential equation itself into the loss function, enforcing physical consistency during training. These frameworks collectively illustrate the unifying philosophy of SciML: merging the interpretability and structure of physical models with the representational flexibility of neural networks.

\paragraph{PIML for holographic duality.}
In recent years, the framework of PIML has been extended to theoretical high-energy physics, particularly to the study of holographic duality: also known as the AdS/CFT correspondence or gauge-gravity duality~\cite{Witten:1998zw,Gubser:1998bc,Maldacena:1997re}. Originally formulated in string theory, this correspondence provides a tractable theoretical methods for strongly interacting quantum systems and quantum gravity and black holes~\cite{Zaanen:2015oix,Hartnoll:2016apf,Ryu:2006bv,Rangamani_2017}.

A pioneering step in connecting deep learning (DL) with holography was made by Hashimoto, Sugishita, Tanaka, and Tomiya~\cite{Hashimoto:2018ftp}, who demonstrated how a bulk geometry can be reconstructed directly from boundary data using a neural network. Their approach, sometimes referred to as the AdS/DL correspondence, showed that once trained on appropriate data, a neural network can infer the emergent bulk metric consistent with the holographic duality framework.\footnote{The conceptual connection between holography and DL had been explored earlier in~\cite{You:2017guh,Gan:2017nyt,Lee:2017skk}. For an accessible introduction to the concept of AdS/DL correspondence, see~\cite{Song:2020agw}.}

This approach effectively addresses an inverse problem: the reconstruction of bulk spacetime from boundary observables, a key challenge in holographic duality. Such bulk reconstruction not only deepens our understanding of the correspondence itself but also provides a computational strategy to infer gravitational duals of strongly coupled quantum systems. The resulting algorithms may, in the future, be implemented in laboratory contexts, offering a data-driven route to explore emergent spacetime phenomena.

Deep learning techniques (Neural ODEs and PINNs) have been applied across a broad spectrum of holographic models, leveraging diverse boundary data from QCD and condensed matter physics. Examples include extracting the magnetization curve in strongly correlated systems~\cite{Hashimoto:2018ftp}, lattice QCD data for the chiral condensate~\cite{Hashimoto:2018bnb,Hashimoto:2020jug}, hadron and meson spectra~\cite{Akutagawa:2020yeo,Hashimoto:2021ihd,Hashimoto:2022eij,Mansouri:2024uwc,Luo:2024iwf}, and transport coefficients such as shear viscosity~\cite{Yan:2020wcd,Gu:2024lrz}. Other studies have explored the QCD equation of state~\cite{Chen:2024ckb,Bea:2024xgv}, phase diagrams of QCD and high-$T_c$ superconductors~\cite{Cai:2024eqa,Kim:2024car}, and optical or anomalous transport phenomena~\cite{Li:2022zjc,Ahn:2024gjf,Ahn:2025tjp}. Further applications include minimal-surface problems in curved geometries~\cite{Hashimoto:2025zmi,Landgren:2025ehb}, systems defined on compact spaces~\cite{Hashimoto:2024yev}, and studies of entanglement entropy~\cite{Park:2022fqy,Park:2023slm,Ahn:2024jkk,Deb:2025kwx,Kim:2025aln}. Collectively, these works illustrate the versatility of DL in capturing the structure of holographic models across diverse physical systems.

\paragraph{Motivation of this paper.}
The equations of motion underlying these holographic models are typically non-linear second-order ordinary differential equations: canonical examples in high-energy theory. Such systems serve as natural benchmarks for applying PINNs and NODEs, both of which are designed to incorporate physical laws into neural architectures.

In this work, we demonstrate the application of Neural ODEs and PINNs to representative inverse problems in holography, focusing on two prototypical cases from applied holography: the QCD equation of state and the $T$-linear resistivity problem. Our objective is to provide a systematic and accessible framework for utilizing neural networks in bulk reconstruction, serving as \textit{a comprehensive and practical reference for researchers entering this area}.

Moreover, we show that such typical holographic problems admit a direct analogy with classical mechanics, allowing the reformulation of gravitational dynamics as mechanical motion in an effective potential. This correspondence offers both conceptual clarity and computational simplicity, extending the reach of PIML techniques beyond holography.\footnote{For further practical guidance and implementation details, including shared hands-on code, see the follow-up manuscript~\cite{WIPHSJ}.}

Finally, the methodologies demonstrated here are broadly applicable beyond high-energy theory. The same principles can be adapted to boundary value problems in mathematics, engineering, and the natural sciences, particularly in systems involving singularities, nonlinearity, or moving boundaries: domains where traditional numerical methods often struggle.

This paper is structured as follows.
Section \ref{sec2} presents a quick overview of the PIML methodologies employed in this work, focusing on neural ordinary differential equations and physics-informed neural networks.
In Section \ref{sec31}, we apply these methods to representative inverse problems in holography, specifically the QCD equation of state and the $T$-linear resistivity problem.
Section \ref{sec32} explores the direct analogy between holographic systems and classical mechanics, illustrating how certain holographic equations can be reformulated as inverse mechanical problems with frictional forces. This section also introduces Kolmogorov-Arnold Networks (KANs) as an alternative approach to neural network-based PIML.
Finally, Section \ref{sec4} concludes with a summary of results and perspectives for future research.

%
\section{Physics-informed machine learning: a method}\label{sec2}
In this section, we present the physics-informed machine learning (PIML) techniques employed in this study, specifically the Neural Ordinary Differential Equations (Neural ODEs) and Physics-Informed Neural Networks (PINNs). Both frameworks serve a dual purpose: they function as numerical solvers for ordinary differential equations to model system dynamics, and as inverse problem solvers to learn the underlying physical parameters, such as the effective potentials, from the observed data.

\subsection{Neural ordinary differential equations}\label{sec:2.1}
We begin by reviewing the concept of Neural ODEs~\cite{Chen:2018wjc,Dupont:2019aa,Massaroli:2020aa,Yan:2019aa}. Consider the second-order ordinary differential equation:
\begin{equation}\label{PIMLODEEQ}
\mathcal{G}(z,\, \phi,\, \partial_z\phi,\, \partial_z^2\phi;\, V(\phi)) = 0 \,, \quad z \in [0\,,1] \,, 
\end{equation}
where $V(\phi)$ represents the potential of $\phi$.\footnote{The original formulation of Neural ODEs involved first-order ODEs~\cite{Chen:2018wjc,Dupont:2019aa,Massaroli:2020aa,Yan:2019aa}. For a detailed algorithm within the context of standard ODE solvers and shooting methods, refer to \cite{Ahn:2024gjf}.} The standard approach to solving this equation involves specifying the potential $V(\phi)$ beforehand and solving for $\phi(z)$:
\begin{equation}\label{PIMLODESolver}
    \phi(z) = \textrm{ODE\;Solver}\left[\mathcal{G} ;\, \{\phi_{(1)}\,, \phi_{\text{(2)}}\};\, V(\phi) \right] \,,
\end{equation}
where $\phi_{(1)}$ and $\phi_{\text{(2)}}$ represent the two free parameters used to solve the equation.

The central idea behind Neural ODEs is to solve the inverse problem: recovering the potential $V(\phi)$ from the given data $\phi(z)$. This is achieved by replacing the arbitrary potential $V(\phi)$ with a deep neural network $\mathcal{D}$, which is expressed as
\begin{equation}\label{PIMLDNN}
    V(\phi) := \mathcal{D} = W_M\cdot \sigma[\cdots\sigma\{W_2\cdot\sigma(W_1 \phi+b_1)+b_2\}\cdots]+b_M \,,
\end{equation}
and solving the ODE equation using the standard ODE solver mentioned above. Note that adaptive time-stepping solvers, such as the Runge-Kutta method of order 4(5), are typically employed in Neural ODE applications to solve the ODEs numerically. This setup can be compared to a general deep neural network architecture \eqref{DNN2} by treating the input layer of single neuron $\phi$ and the output layer as $V(\phi)$.

\paragraph{Data loss function.}
In practice, the potential $V(\phi)$ cannot be arbitrarily fixed from boundary conditions on $\phi$. Instead, the boundary conditions depend on the specific input data the researcher wishes to impose. These boundary conditions are incorporated into the Neural ODE framework via the \textit{Data Loss Function}.

To elaborate, in the examples considered in this manuscript, the ODEs typically exhibit $N$ degrees of freedom (e.g., $N=2$ for the equation \eqref{PIMLODEEQ}). However, by imposing minimal boundary conditions (such as asymptotic AdS boundary conditions), we are left with either one or two free parameters which is equivalent to the number of data loss function. In the case of holographic QCD, for example, we use a single data loss function (the QCD equation of state) to determine a single potential. In contrast, for the holographic condensed matter theory example, two data loss functions, namely the $T$-linear resistivity and linear specific heat, are used to identify two potentials.

Let us illustrate this setup using the simple example from equation \eqref{PIMLODESolver}. We first fix the boundary condition for $\phi_{(1)}$, for instance
\begin{align}\label{BCR}
\begin{split}
\text{Boundary Condition:} \quad \phi(0) =0  \quad\text{for}\quad  \phi_{(1)} \,.
\end{split}
\end{align}
Next, we apply the data loss function to the remaining free parameter, $\phi_{(2)}$, by imposing the relationship $\phi'(0)$ and $\phi(1)$
\begin{align}\label{BCR2}
\begin{split}
\text{Data Loss Function:} \quad  \{ \phi'(0) \,, \phi(1) \}  &\quad\text{for}\quad  \phi_{(2)} \,.
\end{split}
\end{align}
In this example, $\{ \phi'(0) \,, \phi(1) \}$ denotes a certain relationship between $\phi'(0)$ and $\phi(1)$, a typical scenario in classical mechanics problems, as illustrated in Fig. \ref{fig:EoS_conceptual_data}. Nevertheless, the data loss function can be chosen arbitrarily (for instance, one could also examine other relationships like between $\phi(1)$ and $\phi'(1)$). In essence, given the boundary condition and data loss function, one can ultimately determine the potential $V(\phi)$.

For a more precise formulation, the data loss can be expressed as
\begin{equation}\label{DATALOSSS}
    L_{\text{Data}} = \frac{1}{N}\sum_{i=0}^{N-1} | \phi_{\text{data}}^{(i)}(1) - \phi_{\text{trained}}^{(i)}(1) |  \,,
\end{equation}
where $\phi_{\text{data}}^{(i)}(1)$ represents the $i$-th input data point (e.g., $i\in[0,20]$ in Fig. \ref{fig:EoS_conceptual_data}), and $\phi_{\text{trained}}^{(i)}(1)$ is the value obtained by solving the ODEs with the trained potential $V(\phi)$.
\begin{figure}
    \centering
    \includegraphics[width=0.35\linewidth]{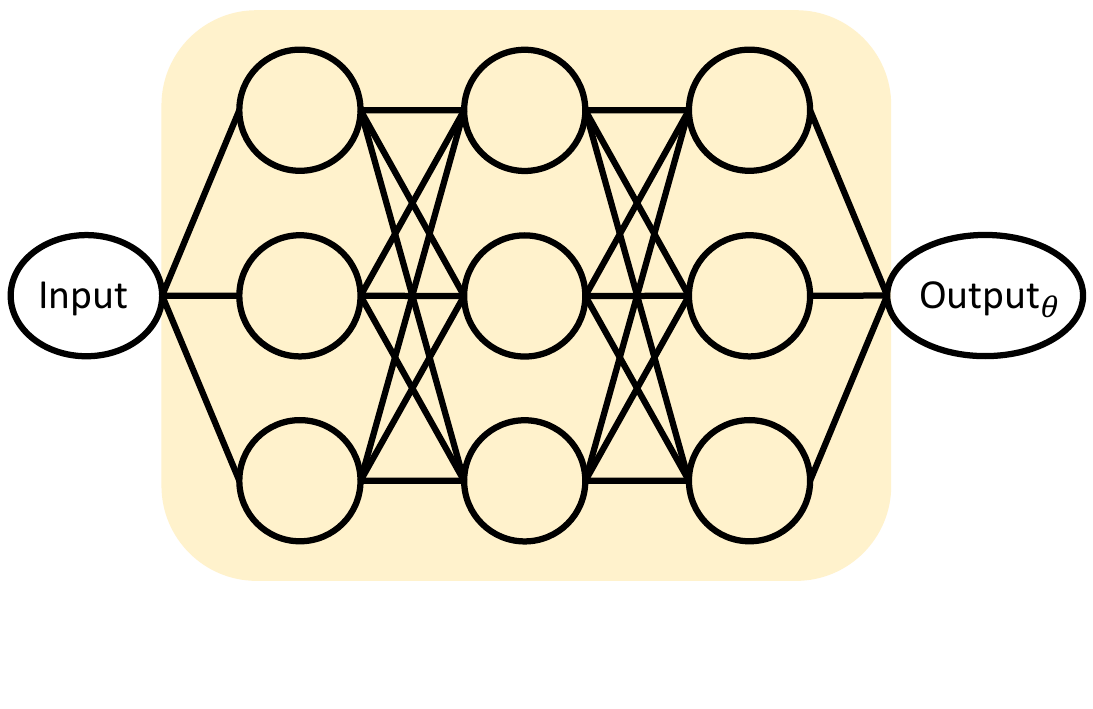}
    \includegraphics[width=0.6\linewidth]{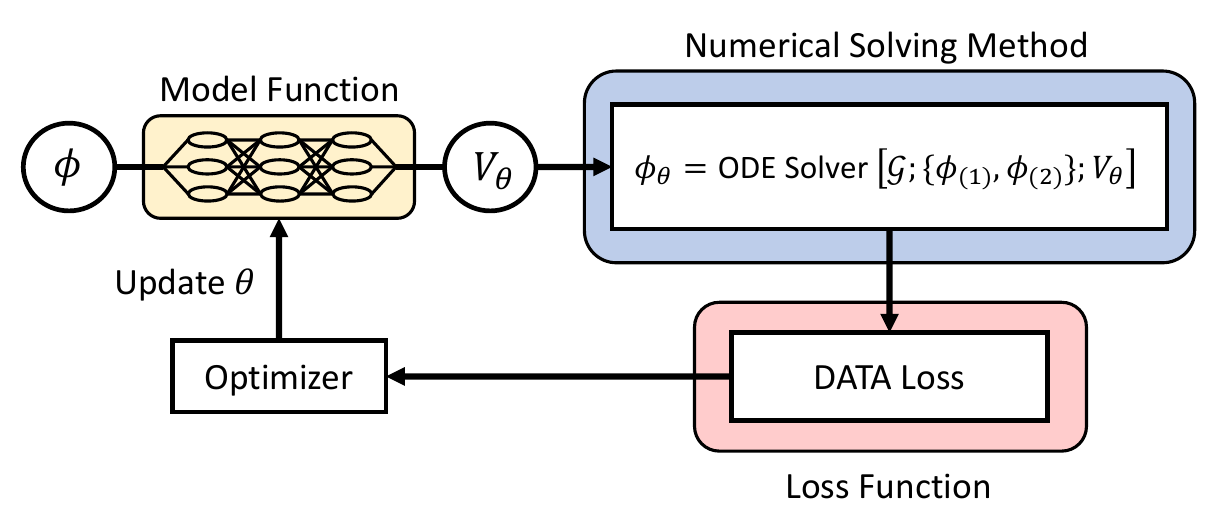}
    \caption{Schematic representation of the model architecture (left) and training workflow of the neural ODE framework (right). The deep neural network, parameterized by $\theta$, predicts the model function $V(\phi)$, which is used in the ODE solver to compute $\phi(z)$. The output is compared with data to form the loss, and the parameters $\theta$ are updated via optimization.}\label{fig:neuralODEdiagram}
\end{figure}
For a schematic overview of Neural ODEs, see Fig. \ref{fig:neuralODEdiagram}.

In this manuscript, in general, \textit{the data loss function refers to the relationship between the field variables}, such as $\{\phi(1)\,, \psi(1)\}$, when solving coupled equations of motion. We will explicitly define the data loss function used for each inverse problem of interest.

\subsection{Physics-informed neural networks}\label{sec:2.2}
\begin{figure}[]
  \centering
     {\includegraphics[width=0.6\linewidth]{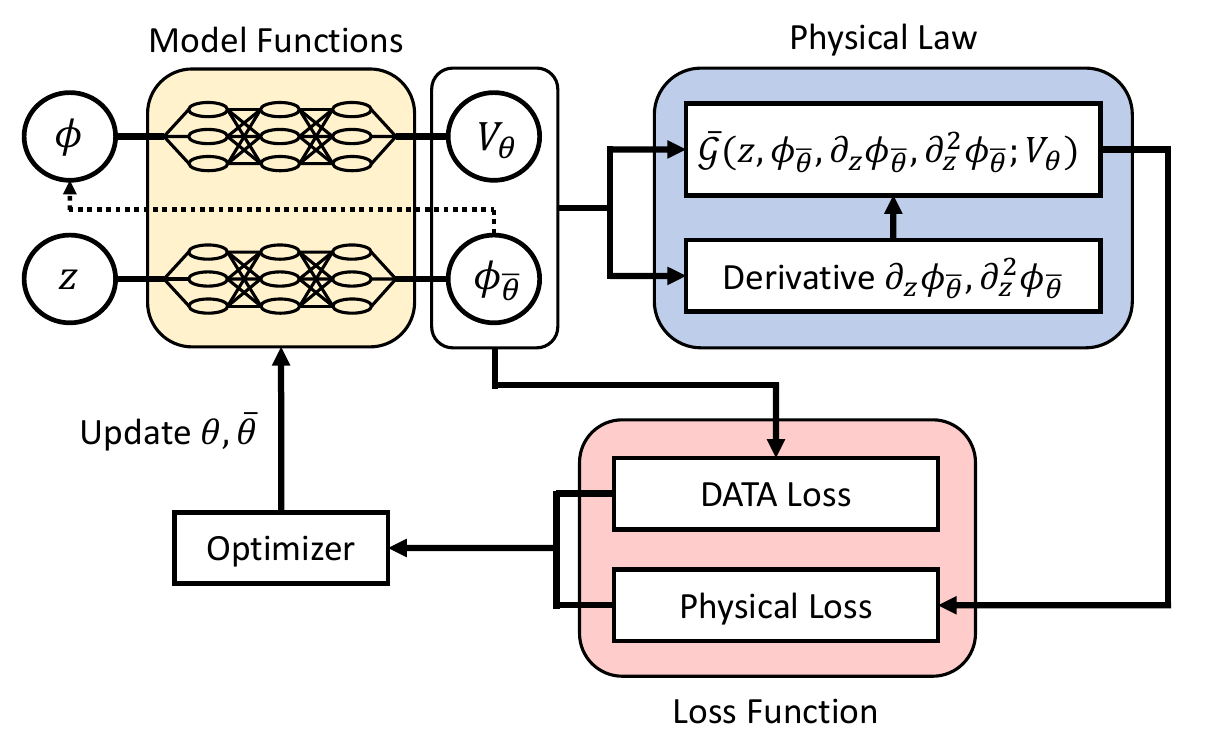} \label{}}
 \caption{Schematic illustration of the PINN framework. The solution $\phi(z)$ and the potential $V(\phi)$ are represented by the deep neural networks parameterized by $\theta$ and $\bar{\theta}$. The network outputs and their derivatives enter the physical loss, which enforces the equations of motion.}\label{SKETCHFIG}
\end{figure}

We now review PINNs~\cite{RAISSI2019686,HBWJ2023,Raissi:2017aa,Raissi:2017ab}. Conceptually, PINNs differ from Neural ODEs in two essential ways. First, in PINNs both the potential $V(\phi)$ and the field $\phi$ are replaced by neural networks, denoted collectively by $\mathcal{D}$'s. The corresponding ODE system can then be rewritten as
\begin{equation}\label{PIMLDNNEQ}
\bar{\mathcal{G}} := \mathcal{G}(z,\, \mathcal{D}_\phi,\, \partial_z\mathcal{D}_\phi,\, \partial_z^2\mathcal{D}_\phi;\, \mathcal{D}_V)  \,,
\end{equation}
which is generally non-vanishing, $\bar{\mathcal{G}}\neq0$, during training. Second, the loss function in PINNs is composed of two parts: the data loss introduced earlier \eqref{DATALOSSS} and the \textit{physical loss}, defined as the residual of the ODEs,
\begin{align}\label{PINNLossEom}
    L_{\text{Physical}} := \frac{1}{N_z}\sum_z \big| \bar{\mathcal{G}} \big|  \,,
\end{align}
so that the total loss reads
\begin{equation}
    L = L_{\text{Data}} + L_{\text{Physical}} \,,
\end{equation}
supplemented with the boundary conditions in \eqref{BCR}. Training the PINN amounts to minimizing this total loss, leading to an approximate satisfaction of the equations of motion, $\bar{\mathcal{G}}\approx0$. A schematic representation of the PINN architecture is shown in Fig. \ref{SKETCHFIG}.

PINNs offer several advantages over Neural ODEs. Since they do not rely on classical ODE solvers, they often converge more rapidly to both the field profile and the potential. They are also particularly effective when the ODEs have complicated functional structure like coupled ODEs, where traditional solvers tend to struggle. Moreover, because both the field and the potential are represented by neural networks, their functional forms are completely flexible and can capture highly nontrivial behavior.

We conclude this section with a few technical remarks. First, boundary conditions such as \eqref{BCR} may be imposed either as an additional loss term (soft constraint) or encoded directly through an ansatz for the fields and potentials (hard constraint). In this work, we implement the initial condition as a hard constraint, which facilitates stable convergence.

Second, in PINNs the neural network for the potential $V(\phi)$ must yield a consistent function across all input data points $\phi_{\text{data}}^{(i)}(1)$. In contrast, another neural network representing $\phi(z)$ apparently has multiple outputs, one for each data point. This additional structure is required because, unlike in Neural ODEs, the field itself is parameterized by a neural network. Practically, this is implemented by assigning multiple neurons in the output layer of the field network, with each neuron corresponding to a different $\phi_{\text{data}}^{(i)}(1)$.

Last but not least, the choice of loss function can vary: standard options include the Mean Absolute Error (L1 loss) and Mean Squared Error (L2 loss), or more adaptive schemes designed to balance small and large errors. In this manuscript, we employ the L1 loss.

%
\section{Neural network approaches to inverse problems: handful examples}\label{}
Implementing the PIML methodologies described in Section \ref{sec2}, including Neural ODEs and PINNs, we address inverse problems in holography, specifically examining two representative cases from applied holography: the QCD equation of state and the $T$-linear resistivity of holographic strange metals. Furthermore, we demonstrate that these typical holographic problems can be directly analogized to classical mechanics, enabling the reformulation of gravitational dynamics as mechanical motion within an effective potential.

\subsection{Holographic QCD and condensed matter}\label{sec31}

\subsubsection{QCD equation of state}\label{sec:QCD}
The AdS/CFT correspondence has long been hoped to provide insights into the complex problem of quantum chromodynamics (QCD), as initially articulated in early works such as \cite{Witten:1998zw}. Since then, a substantial and diverse body of literature has developed as the idea of ``AdS/QCD". Key references in this field include \cite{Csaki:1998qr,Polchinski:2001tt,Sakai:2004cn,Erlich:2005qh}, and comprehensive reviews can be found in \cite{CasalderreySolana:2011us,Ammon:2015wua,Natsuume:2014sfa}.

\paragraph{Equation of state (or speed of sound).}
A fundamental thermodynamic question arises when considering a holographic gravity dual of QCD: whether the equation of state accurately reflects the physical properties of QCD. The AdS/CFT framework has motivated the use of thermal $\mathcal{N}=4$ super Yang-Mills (SYM) theory to model QCD at finite temperatures. However, a limitation of SYM theory is its conformal invariance, which implies that the speed of sound ($C_S$) remains constant at $1/\sqrt{3}$ across all temperatures. In contrast, QCD only exhibits conformal behavior at high temperatures. To better capture QCD's behavior over a broader temperature range, gravity duals of gauge theories that break conformal invariance are considered.

The minimal gravity action describing Lorentz-invariant, non-conformal theories is given by:
\begin{equation}\label{QCDaction}
S = \int \dd^5x \sqrt{-g}  \left[ R - \frac{1}{2} (\nabla \Phi)^2 - V(\Phi) \right]  \,,\end{equation}
Here, $V(\Phi)$ remains arbitrary. For simplicity, we focus on the homogeneous ansatz:
\begin{equation}\label{METRIC}
\dd s^2 = \frac{1}{z^2}\Big( -f(z)\,\dd t^2 + \frac{1}{f(z)}\,\dd z^2 + h(z) \sum_i \dd x_i^2 \Big) \,,
\end{equation}
where $z=1$ corresponds to the horizon radius, and $z=0$ represents the AdS boundary. The bulk equations of motion derived from the action \eqref{QCDaction}  read
\begin{equation}\label{EOSEOM}
\begin{split}
&2 h(z) \big(V(\Phi(z)) - 3 z f'(z)\big)
+ 3 z^2 f'(z) h'(z)
\\
&\qquad + f(z) \big[h(z) \big(24 + z^2 \Phi'(z)^2 \big)
+ 6 z \big(-3 h'(z) + z h''(z)\big)\big] = 0, 
\\[4pt]
& f(z)\big(-3 z^2 h'(z)^2
+ h(z)^2 \big(24 + z^2 \Phi'(z)^2 \big)\big)
\\
&\qquad + 2 h(z) \big(3 z^2 f'(z) h'(z)
+ h(z)\big(V(\Phi(z))
+ 3 z \big(-4 f'(z) + z f''(z)\big)\big)\big) = 0,
\\[4pt]
& h(z) \big(2 h(z) \big(V(\Phi(z)) - 3 z f'(z)\big)
+ 3 z^2 f'(z) h'(z)\big)
\\
&\qquad + f(z) \big(-18 z h(z) h'(z)
+ 3 z^2 h'(z)^2
+ h(z)^2 \big(24 - z^2 \Phi'(z)^2\big)\big) = 0,
\\[4pt]
& 3 z^2 f(z) h'(z) \Phi'(z)
\\
&\qquad + 2 h(z)\big(-V'(\Phi(z))
+ z\big((-3 f(z) + z f'(z)) \Phi'(z)
+ z f(z) \Phi''(z)\big)\big) = 0 \;.
\end{split}
\end{equation}

At small $\Phi$, and using the asymptotic AdS boundary conditions $f(0)=h(0)=1$, the potential $V(\Phi)$ takes the form
\begin{equation}\label{VASMY}
    V(\Phi) = -12 + \frac{1}{2}m^2\Phi^2 + \mathcal{O}(\Phi^3)  \,.
\end{equation}
An asymptotically AdS spacetime corresponds to conformal invariance in the UV limit. Gravity backgrounds constructed from potentials that satisfy this form are dual to deformations of a conformal field theory
 \begin{equation}\label{FTLagrangian}
   {\cal L} = {\cal L}_{\rm CFT} + \Lambda_\Phi^{4-\Delta} {\cal O}_{\Phi} \,,
 \end{equation}
where $\Lambda_\Phi$ is the energy scale of the deformation and $\Delta$ is the dimension of the operator ${\cal O}_\Phi$. The AdS/CFT dictionary relates $\Delta$ to the mass $m$ of the scalar field as
 \begin{equation}\label{DELm}
   \Delta(\Delta - 4) = m^2 \,.
 \end{equation}

Within the given background \eqref{METRIC}, the black hole thermodynamic quantities, evaluated at the event horizon, are
\begin{equation}
    s = 4\pi h(1)^{\frac{3}{2}}  \,, \quad T = - \frac{f'(1)}{4\pi} \,.
\end{equation}
where $s$ is the thermal entropy and $T$ Hawking temperature. Then, the sound speed squared, $C_S^2$ , is calculated from
\begin{equation}
    C_S^2 = \frac{d\,\log T}{d\,\log s} \,,
\end{equation}
where we exclude the chemical potential for baryon number for simplicity.

\begin{figure}[]
  \centering
     \includegraphics[width=\linewidth]{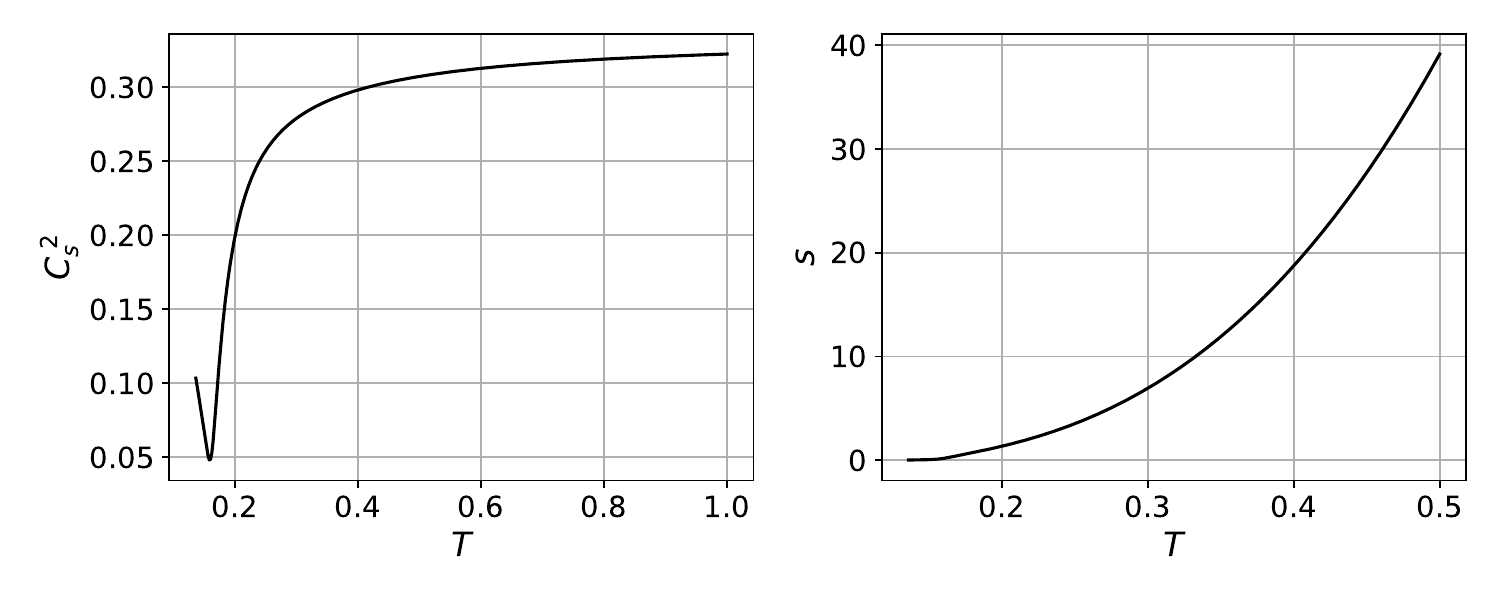}
 \caption{Speed of sound $C_S^2$ (left) and equation of state $s(T)$ (right) for the potential $V(\phi)$ in \eqref{FirstPotential}. The critical temperature is around $T_c\approx0.18$, and at high temperatures, $C_S^2$ approaches $1/3\approx 0.33$, as expected for conformal behavior.}\label{FIG:EoS_data}
\end{figure}

Remarkably, Gubser, Nellore, Pufu, and Rocha \cite{Gubser:2008yx,Gubser:2008ny} found a surprisingly simple potential $V(\Phi)$ that approximates the lattice data for QCD
\begin{equation}\label{FirstPotential}
   V(\Phi) = -12 \cosh(\gamma \, \Phi) + b \, \Phi^2 \,, \quad \text{with} \quad \gamma = 0.606 \,,\,\, b = 2.057 \,.
 \end{equation}
This potential reproduces the sound speed squared vs. temperature derived from lattice QCD for $(2+1)$-flavor QCD \cite{Cheng:2007jq}: see Fig. \ref{FIG:EoS_data}.\footnote{The equation of state derived from \eqref{FirstPotential} closely matches the quasiparticle model proposed in \cite{Bluhm:2007nu}, which is constructed through a chiral extrapolation of lattice QCD data.} The equation of state from this potential exhibits a crossover rather than a sharp phase transition, with the critical temperature $T_c$.

\paragraph{PINNs with QCD equation of state.}
We now apply PINNs to solve the inverse problem using the QCD equation of state data shown in the figure above, which corresponds to the data loss function \eqref{DATALOSSS}: $\{T\,,s\} \approx \{f'(1)\,, h(1)\}$ and the physical loss function \eqref{PINNLossEom} built from \eqref{EOSEOM}. As demonstrated in Sec. \ref{sec2}, we use the hard-constraint scheme for equation of state data here.

Specifically, we use the data for $C_S^2$ to identify the bulk fields (metric $g_{\mu \nu}(z)$ and matter field $\phi(z)$) and the potential $V(\phi)$. Recent work has also applied PIML to the similar problem of using thermodynamic quantities, e.g., \cite{Chen:2024ckb,Bea:2024xgv}, but we find PINNs to be more stable for this particular equation of state, as compared to methods like Neural ODE.

To compute the equation of state as a function of temperature, we keep $\lambda_{\Phi}$ in \eqref{FTLagrangian} fixed, setting $\lambda_{\Phi}=1$, which corresponds to a fixed scalar field source term~\cite{Gubser:2008yx,Gubser:2008ny}. For convenience, we redefine the dilaton field as
\begin{equation}
    \Phi(z) := z^{\Delta}\phi(z) \,,
\end{equation}
where $\Delta$ is given in \eqref{DELm}, with $\phi(z) \approx \phi_0 + \mathcal{O}(z)$ near the AdS boundary and $\phi(0)=\phi_0=1$.

Thus, all the constraints on the fields can be summarized as 
\begin{equation}\label{EoS_constraint}
\begin{split}
    f(0)=1 \;, \quad h(0)&=1 \;, \quad \phi(0)=1  \;, \\
    f(1)=0 \;, \quad& T = - \frac{f'(1)}{4\pi} \;.
\end{split}
\end{equation}
including asymptotic AdS boundary and black hole conditions: these five constraints arise because one of the bulk equations of motion in \eqref{EOSEOM} is redundant. These constraints are embedded into the neural network ansatz
\begin{equation}
\begin{split}
    f(z) &:= (1-z) - z(1-z)(1 - 4\pi T) + z(1-z)^2 \mathcal{D}_f(z) \;, \\
    h(z) &:= (1-z) + z\,\left(\frac{s}{4\pi}\right)^{\frac{2}{3}} + z(1-z)\mathcal{D}_h(z) \;, \\
    \phi(z) &:= 1 + z\mathcal{D}_\phi(z) \,,
\end{split}
\end{equation}
where the neural networks $\mathcal{D}$ have 3 hidden layers with 100 nodes and SoftPlus activation, and the output layer has 21 nodes (one for each temperature data point).

On the other hand, the potential $V(\phi)$ is parameterized as
\begin{equation}
    V(\phi) := -12 + \frac{1}{2}m^2\phi^2 + \phi^3\mathcal{D}_V(\phi) \;  ,
\end{equation}
where $\mathcal{D}_V$ has 3 hidden layers with 20 nodes and SoftPlus activation.

\begin{figure}[]
  \centering
     {\includegraphics[width=0.5\linewidth]{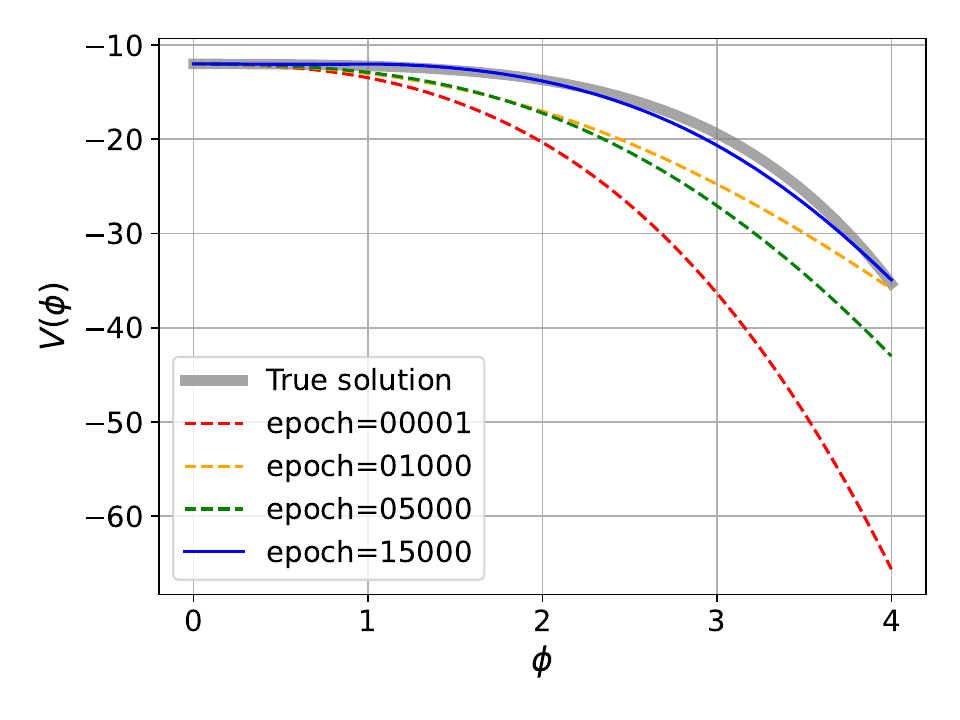} \label{}}
 \caption{The trained potential $V(\phi)$ from the QCD equation of state, Fig. \ref{FIG:EoS_data}, using PINNs. The true solution is \eqref{FirstPotential}.}\label{FIG:EoS_potential}
\end{figure}

\begin{figure}[]
  \centering
     {\includegraphics[width=0.32\linewidth]{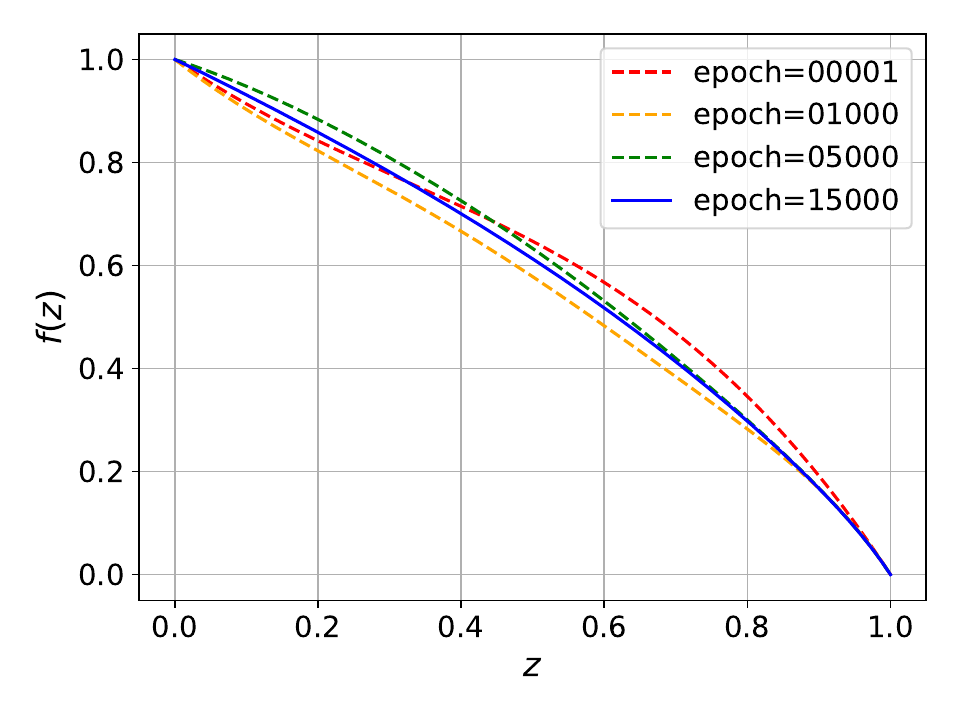} \label{}}
     {\includegraphics[width=0.32\linewidth]{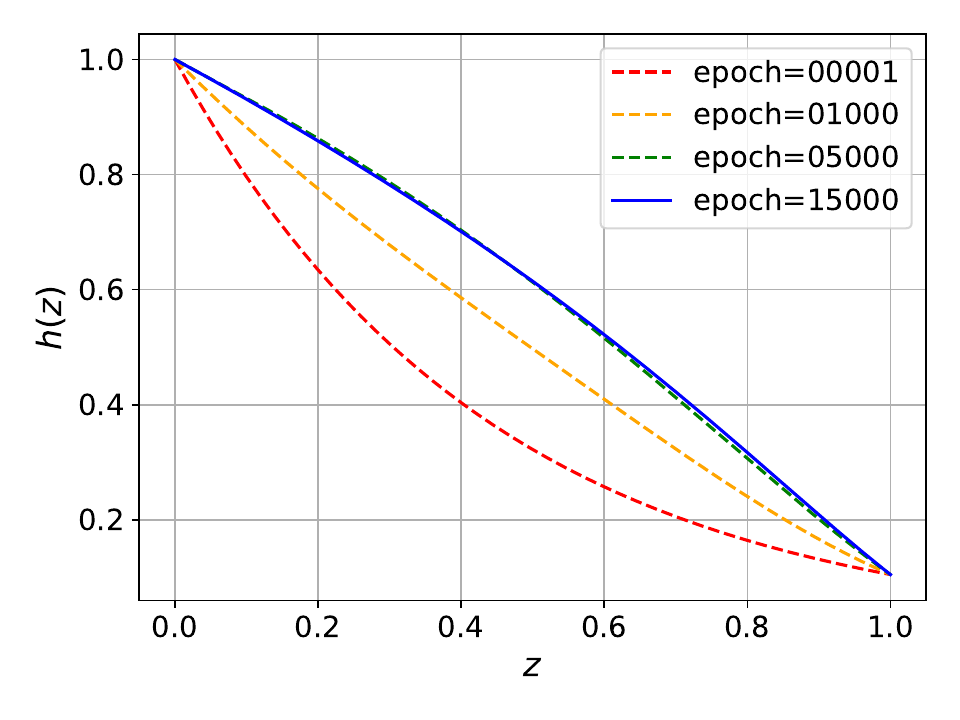} \label{}}
     {\includegraphics[width=0.32\linewidth]{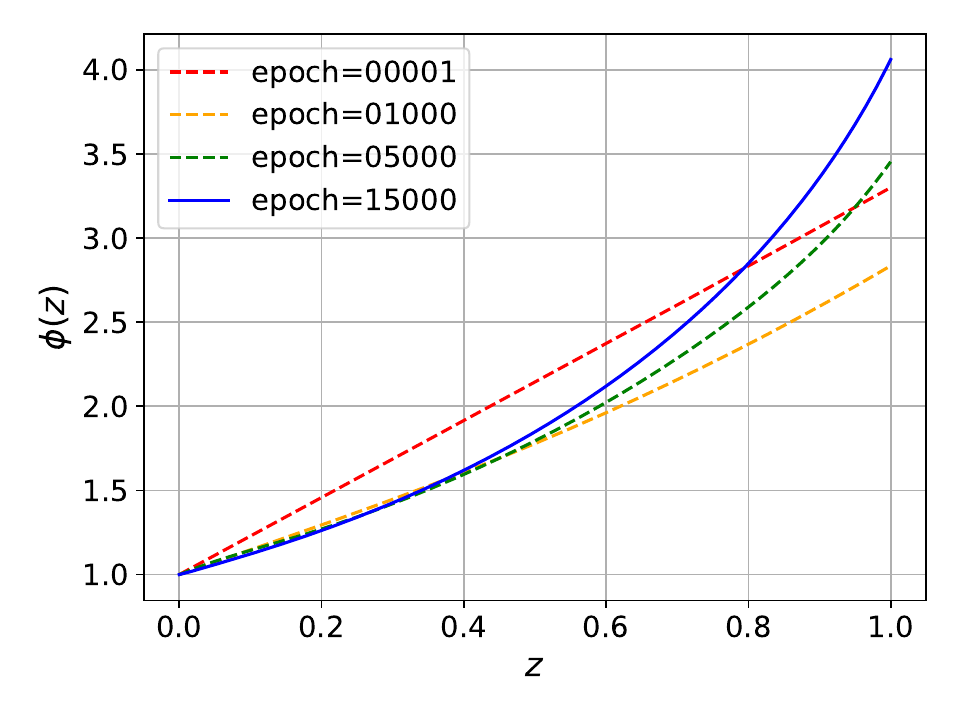} \label{}}
 \caption{The trained bulk fields using PINNs from the QCD equation of state Fig. \ref{FIG:EoS_data}. }\label{FIG:EoS_fields}
\end{figure}

The results from our PINNs are displayed in Fig. \ref{FIG:EoS_potential} and \ref{FIG:EoS_fields}, demonstrating that PINNs can effectively solve the inverse problem and recover the potential $V(\phi)$ \eqref{FirstPotential}.
Our results further indicate that PIML provides a powerful framework for addressing inverse problems in AdS/QCD, enabling the construction of more robust dual gravity models. In our companion work~\cite{wipMLteam}, we will extend this approach to investigate transport properties, such as the shear viscosity-to-entropy density ratio, in strongly coupled quark-gluon plasma~\cite{Kovtun:2004de,Policastro:2001yc}.

\subsubsection{$T$-linear resistivity of strange metals}
The AdS/CFT correspondence is inherently able to study strongly interacting quantum systems, such as QCD via AdS/QCD, and has been extended to investigate strongly correlated condensed matter systems (CMT), known as ``AdS/CMT". This approach allows for the study of quantum critical states of matter in a controlled and theoretically tractable manner~\cite{Zaanen:2015oix, Hartnoll:2016apf, Natsuume:2014sfa}. For a comprehensive recent review, see \cite{Cinti:2025kqh}.

\paragraph{$T$-linear resistivity.}
A key motivation behind AdS/CMT is the study of anomalous transport properties observed in strange metals, particularly the $T$-linear resistivity observed in normal phases of high-$T_c$ superconductors~\cite{Zaanen:2010yk,Varma:1989aa,Hussey__2004}. This resistivity exhibits a universal linear relationship with temperature, expressed as
\begin{equation}
\rho \approx T \,,
\end{equation}
where $\rho$ is the electric DC resistivity, which shows a linear dependence on temperature $T$ in various strange metals, including cuprates, pnictides, and heavy fermion systems. This contrasts sharply with the $T^2$ scaling found in ordinary metals governed by Fermi liquid theory. The universality of this behavior suggests that a deeper understanding of its underlying mechanisms is necessary.

A prominent holographic framework for modeling $T$-linear resistivity was identified by Davison, Schalm, and Zaanen~\cite{Davison:2013txa}, initially proposed by Gubser and Rocha~\cite{Gubser:2009qt}. This simple Einstein-Maxwell-Dilaton-axion theory captures $T$-linear resistivity and the scaling of entropy density with temperature, $s \sim T$, which has been observed in the strange metal phase of cuprates. The model is described by the action
\begin{equation}\label{EMDA}
    S = \int \dd^{4}x  \sqrt{-g}\left( R  - \frac{1}{2}(\partial \phi)^2 + V(\phi) - \frac{Z(\phi)}{4}F^{2} - \frac{1}{2}\sum_{i=1}^{2} (\partial \chi_i)^2 \right) \,,
\end{equation}
with the potential and coupling functions
\begin{equation}\label{GRmodel}
    V(\phi) = 6 \cosh \frac{\phi}{\sqrt{3}} \;, \quad Z(\phi) = e^{\frac{\phi}{\sqrt{3}}} \,.
\end{equation}
This framework is widely recognized as one of the most successful holographic models for describing $T$-linear resistivity and the associated thermodynamic behaviors of strange metals.

The Gubser-Rocha model \eqref{GRmodel} has several advantages: (I) it provides the UV-completed analytic solutions, rare among holographic models at finite temperature~\cite{Gubser:2009qt,Davison:2013txa,Jeong:2018tua,Jeong:2021wiu,Ahn:2023ciq}, (II) It is rooted in string theory, emerging from D3-branes in type IIB string theory ($4+1$ dimensions) and consistent truncation of eleven-dimensional supergravity on AdS$_4 \times S^7$ ($3+1$ dimensions)~\cite{Gubser:2009qt, Cvetic:1999xp}, (III) In IR resistivity analysis, it shows $T$-linear resistivity, linked to semilocal quantum liquids and the AdS$_2$ geometry~\cite{Gouteraux:2014hca}, relevant in condensed matter models like SYK models.

\paragraph{Neural ODEs with $T$-linear resistivity.}
Notably, recent work \cite{Ahn:2025tjp} has shown the utility of PINNs in identifying the unknown potential and coupling in \eqref{EMDA} from diverse $T$-linear resistivity data, where it also re-discovered the Gubser-Rocha model \eqref{GRmodel}. In the current manuscript, we introduce an alternative approach using Neural ODE, which provides a consistent framework for solving inverse problems in holographic condensed matter theories.

We begin with the ansatz for the bulk metric and fields
\begin{align} \label{Ourmodelmetric}
\begin{split}
        \dd s^2 &= \frac{1}{z^2}\left(-f(z) \dd t^2+\frac{1}{f(z)}\dd z^2+h(z)\sum_{i=1}^{2}\dd x_{i}^{2}\right) \;, \\
        \phi &= \phi(z) \,, \qquad A=A_t(z) \dd t \,, \qquad \chi_{i} = \beta x_{i} \,,       
\end{split}
\end{align}
where $\beta$ controls translational symmetry breaking, which is crucial for describing finite resistivity, where we fix $\beta/\mu=10$ for simplicity. The equations of motion derived from the action \eqref{EMDA} give rise to a system of differential equations:
\begin{equation}\label{linearT_eq}
    \begin{aligned}
    & 0 = 2 z^2 f(z) h'(z) \phi'(z) + 2 h(z) V'(\phi(z)) \\ 
    & \qquad + z h(z) \Big(z^3 A_t'(z)^2 Z'( \phi(z) )+2 \left(z f'(z)-2 f(z)\right) \phi '(z) + 2 z f(z) \phi''(z)\Big) \;, \\
    & 0 = A_t''(z)+\frac{A_t'(z) h'(z)}{h(z)}+\frac{A_t'(z) \phi'(z) Z'(\phi (z))}{Z(\phi (z))} \;, \\
    & 0 = \frac{2 h''(z)}{h(z)}-\frac{h'(z)^2}{h(z)^2}+\phi'(z)^2 \;, \\ 
    & 0 =  h(z)^2 \Big(z^4 A_t'(z)^2 Z(\phi (z)) -4 z f'(z)+f(z) \left(12-z^2 \phi '(z)^2\right) - 2 V(\phi(z)) \Big) \\
    & \qquad + 2 z h(z) \Big(\left(z f'(z)-4 f(z)\right) h'(z)+\beta ^2 z \Big) + z^2 f(z) h'(z)^2  \,,
    \end{aligned}
\end{equation}
where we compile the independent equations of motion, which involve seven free constraints. These constraints are applied to enforce the asymptotic AdS boundary conditions, as well as the black hole and regularity conditions at the horizon:
\begin{equation}\label{linearT_BC}
\begin{split}
    f(0) = h(0) = 1 \,, \quad A_t(0) = \mu \,, \quad \phi(0) = 0 \,,  \\
    f(1) = 0 \,, \quad A_t(1) = 0 \,, \quad T = \frac{-f'(1)}{4\pi} \,,
\end{split}
\end{equation}
where $\mu$ is interpreted as the chemical potential on the boundary theories.

\begin{figure}[]
  \centering
     {\includegraphics[width=6.8cm]{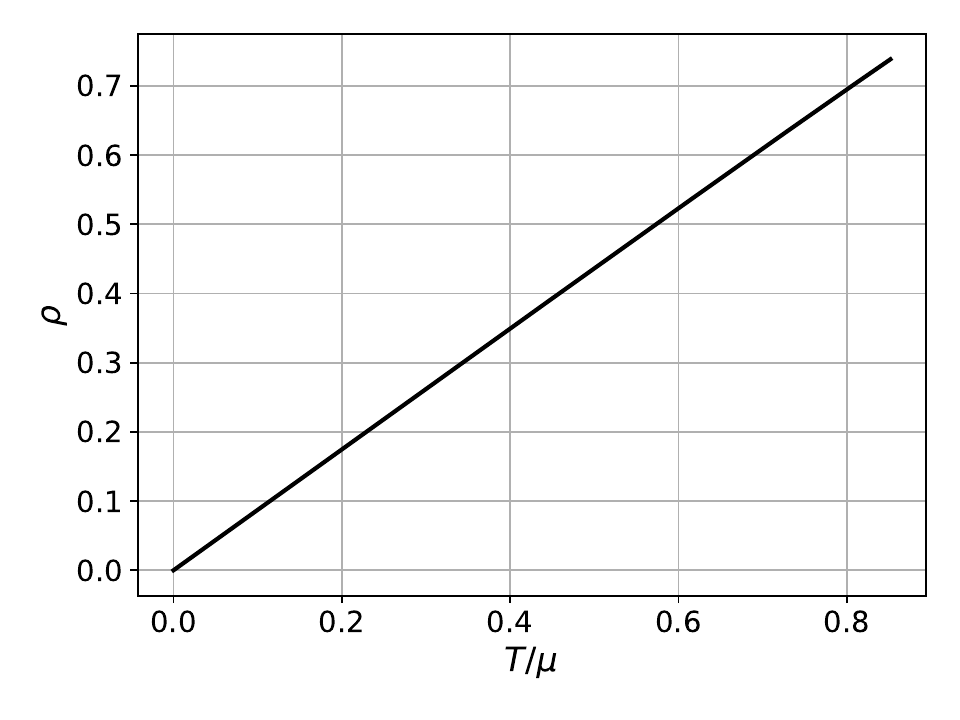} \label{}}
\quad
     {\includegraphics[width=6.8cm]{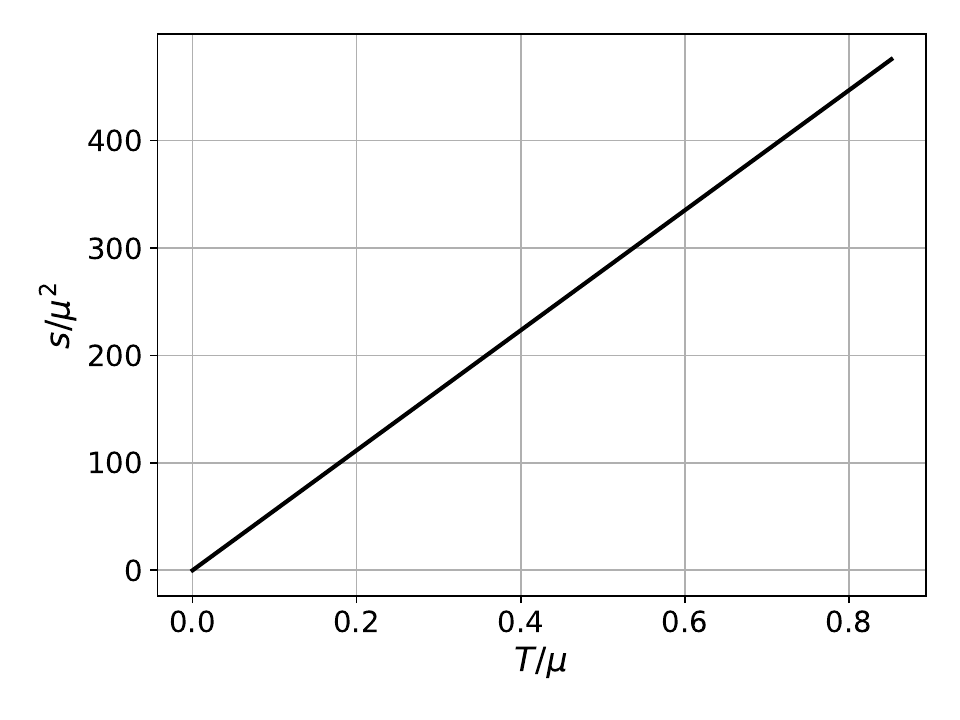} \label{}}
 \caption{$T$-linear resistivity and linear specific heat of Gubser-Rocha model \eqref{GRmodel}.}\label{FIG:linearT_data}
\end{figure}

As an inverse problem, we utilize $T$-linear resistivity and specific heat data to determine the potential and coupling functions. For this purpose, we use the data from the Gubser-Rocha model \eqref{GRmodel}, which is displayed in Fig. \ref{FIG:linearT_data}.
The resistivity and entropy are computed from the horizon formulas
\begin{equation}\label{linearT_Data}
    \rho = \left[ Z(\phi(1))  + \frac{\left[Z(\phi(1)) \, h(1) \, A_t'(1)\right]^2}{\beta^2 \, h(1)} \right]^{-1} \,, \qquad s = 4\pi \, h(1) \,,
\end{equation}
and the corresponding UV boundary conditions for the potential and coupling are given by
\begin{equation}\label{linearT_UV}
    V(\phi) = 6 - \frac{1}{2}m^2 \phi^2 + \cdots \,, \qquad Z(\phi) = 1 + \cdots  \,.
\end{equation}

Finally, to solve our inverse problem, we adopt the following anstaz for the potential and coupling
\begin{align}\label{Ourmodelfields}
\begin{split}
    V(\phi) &:= 6 - \frac{1}{2} m^2 \phi^2 + \phi^3 \mathcal{D}_V(\phi) \,, \\
    Z(\phi) &:= 1 + \phi \mathcal{D}_Z(\phi)  \,,
\end{split}
\end{align}
where $\mathcal{D}$ denote deep neural networks with 5 hidden layers and 20 nodes in each layer. Implementing standard ODE solver (DOPRI 5), we solve the bulk equations of motion \eqref{linearT_eq}, by minimizing the data loss function
\begin{align}\label{LOSS2}
    L_{\text{data}} = \frac{1}{N_{T}}\sum_{T} |\rho(T/\mu) - \rho_{\text{data}}(T/\mu)| \,, \qquad N_T=20 \,,
\end{align}
along with the boundary loss function based on \eqref{linearT_BC} and the linear specific heat condition is also imposed. By solving these equations, we determine the bulk fields, and their corresponding potential and coupling, which can accurately reproduce the $T$-linear resistivity and specific heat.

\begin{figure}[]
  \centering
     {\includegraphics[width=0.4\linewidth]{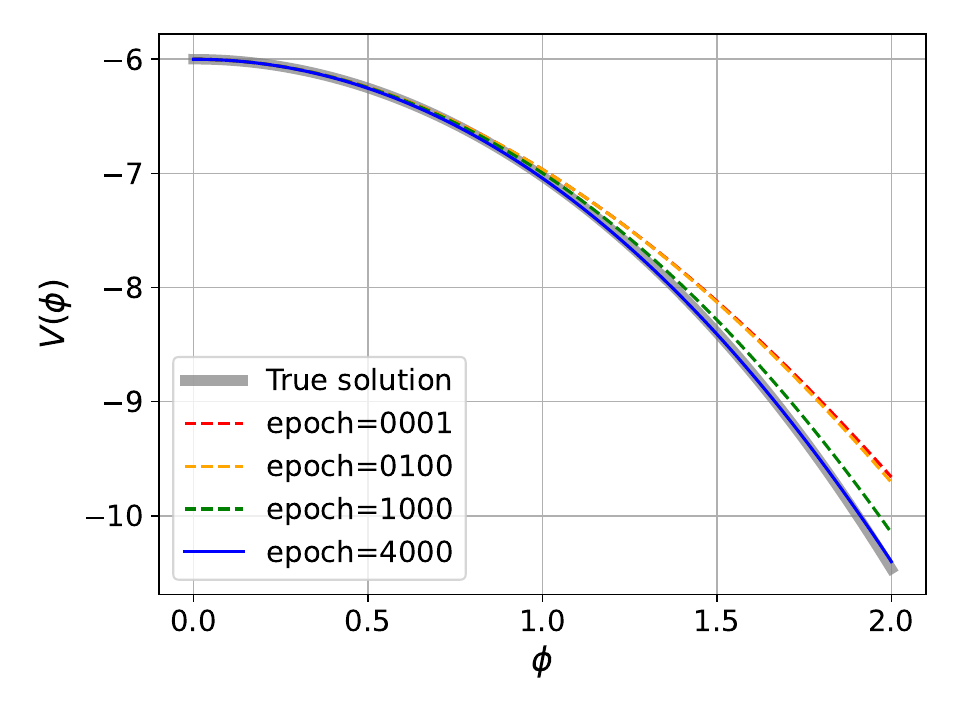} \label{}}
\quad
     {\includegraphics[width=0.4\linewidth]{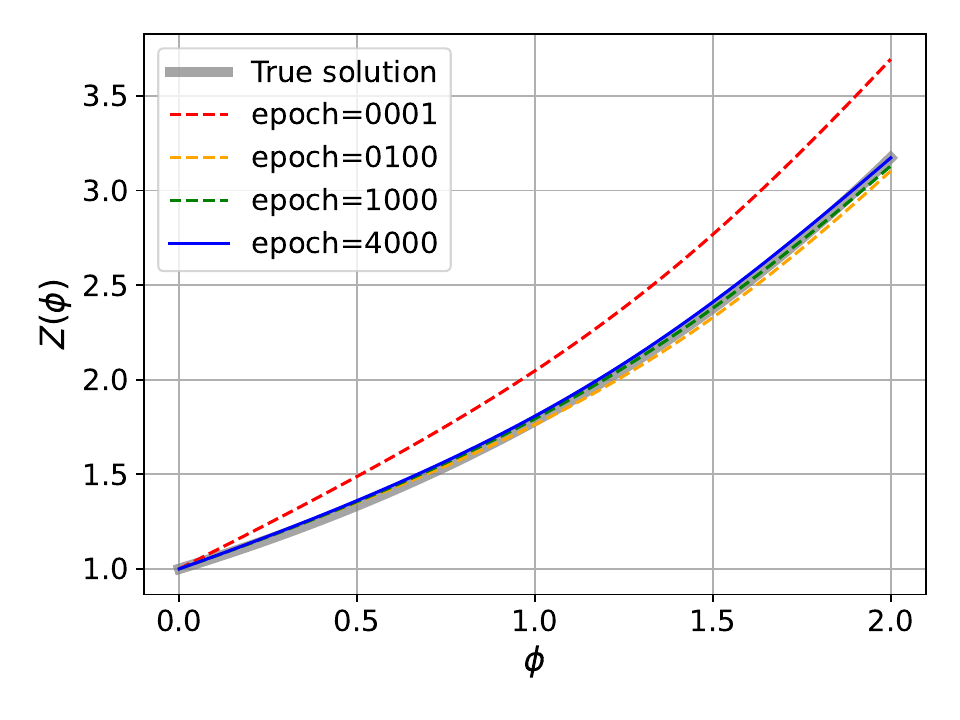} \label{}}
 \caption{The trained potential $V(\phi)$ and coupling $Z(\phi)$ from holographic strange metals, Fig. \ref{FIG:linearT_data}, using Neural ODE. The true solution is \eqref{GRmodel}.}\label{FIG:linearT_result}
\end{figure}

\begin{figure}[]
  \centering
     {\includegraphics[width=0.22\linewidth]{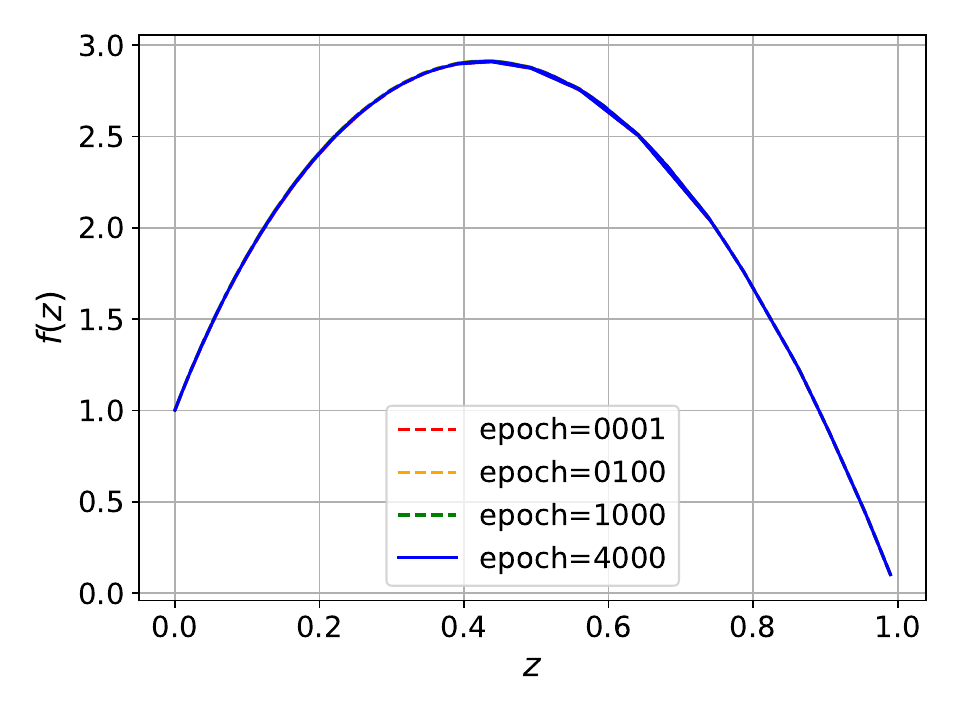} \label{}}
     {\includegraphics[width=0.22\linewidth]{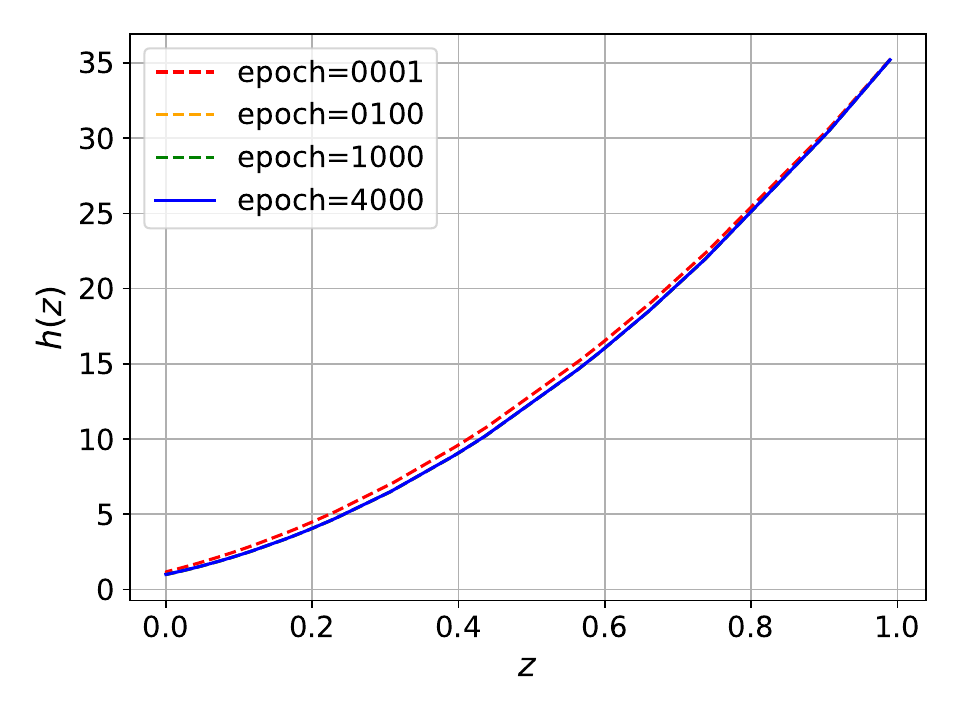} \label{}}
     {\includegraphics[width=0.22\linewidth]{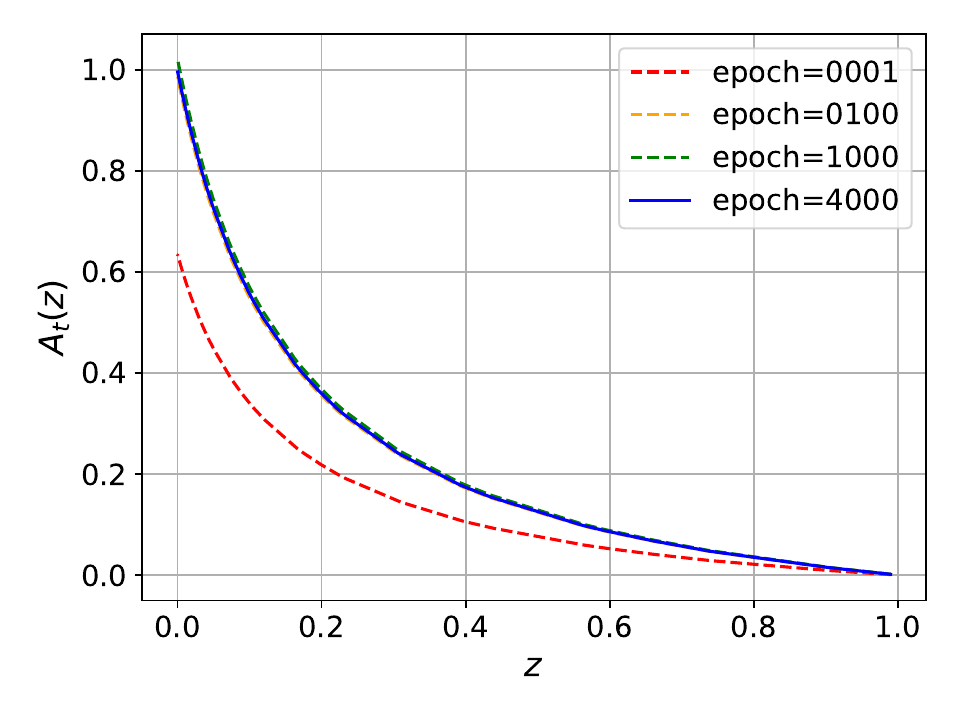} \label{}}
     {\includegraphics[width=0.22\linewidth]{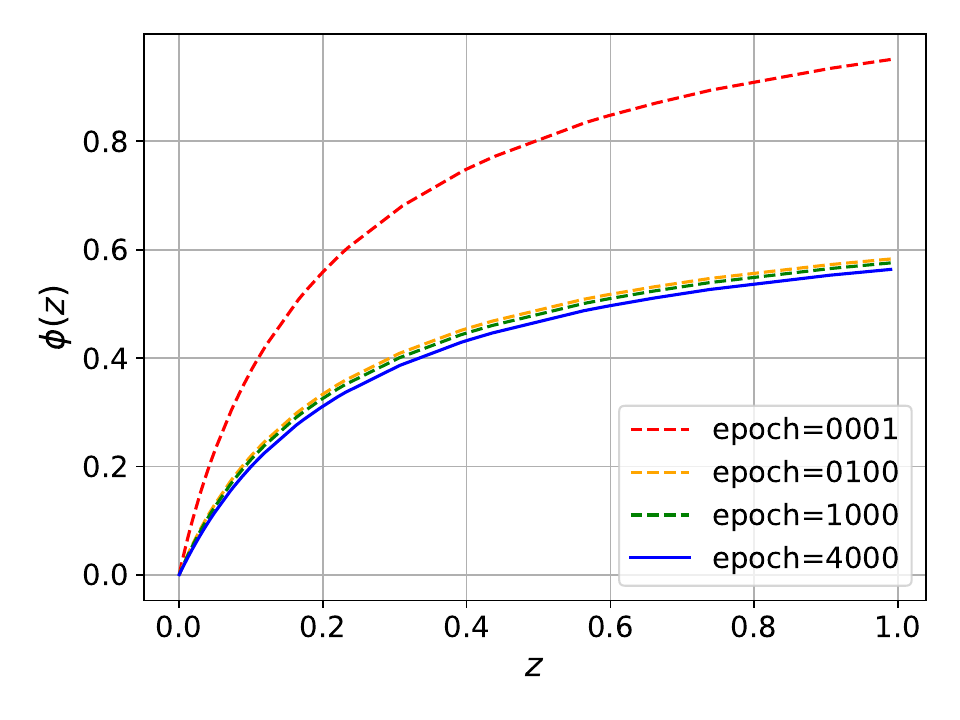} \label{}}
 \caption{The trained bulk fields using Neural ODE from holographic strange metals. }\label{FIG:linearT_fields}
\end{figure}

Our results demonstrate the effectiveness of Neural ODE in solving the inverse problem of identifying the unknown potentials $V(\phi)$ and $Z(\phi)$, with successful reconstruction from the $T$-linear resistivity and specific heat data. The trained potentials and bulk fields obtained using this method are shown in Figs. \ref{FIG:linearT_result} and \ref{FIG:linearT_fields}, where the solutions match the analytic predictions of the Gubser-Rocha model.
In conclusion, our approach with Neural ODE offers a powerful tool for addressing inverse problems in AdS/CMT as well as in AdS/QCD, complementing previous methods such as PINNs. The results provide a consistent framework for constructing more robust dual gravity models and offer promising directions for further research in holographic condensed matter theories.

\subsection{Classical mechanics}\label{sec32}
In this section, we explore how classical mechanics can serve as a simple yet effective analogy for addressing inverse problems in holography. By drawing parallels between mechanical systems and holographic models, we aim to provide a more intuitive understanding of these complex physical phenomena.

\subsubsection{Simple yet effective analogy to holographic problem}\label{sec:analogy}
Consider an object sliding down a frictional incline at an angle $\theta$. This system, governed by Newton's second law, is described by the second-order differential equation
\begin{equation}\label{EOM}
\ddot{x} = g \sin\theta - \frac{F(x)}{m} \,, \quad t \in [0\,,1] \,
\end{equation}
where $F(x)$ represents the frictional force, $x$ is the displacement, and $m$ is the mass of the object. The parameters are fixed as $\theta = 30^\circ$, $m = 1$, and $g = 10$.

The typical setup involves the object starting at the origin with an initial velocity, with motion driven by gravity and opposed by friction
\begin{equation}\label{CM_initial}
x(0) = 0, \qquad \dot{x}(0) = \text{Initial Condition} \;,
\end{equation}
where the initial velocity $x'(0)$ serves as the input parameter. The direct problem in this case is to determine the trajectory $x(t)$ given the frictional force. For instance, one of the non-trivial frictional models is viscous friction, $F(x) \approx \dot{x}$, where the frictional force is proportional to the velocity and opposes the motion. This model is often used in scenarios involving motion through a resistive medium or fluid.

\begin{figure}[]
  \centering
  \includegraphics[width=0.4\textwidth]{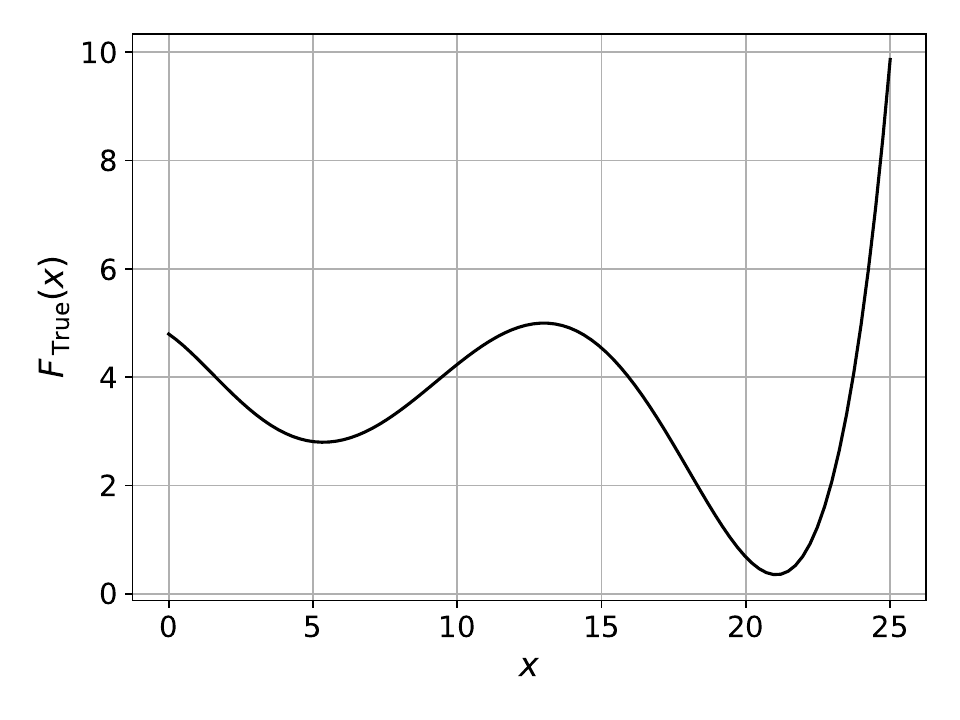}
\qquad
  \includegraphics[width=0.4\textwidth]{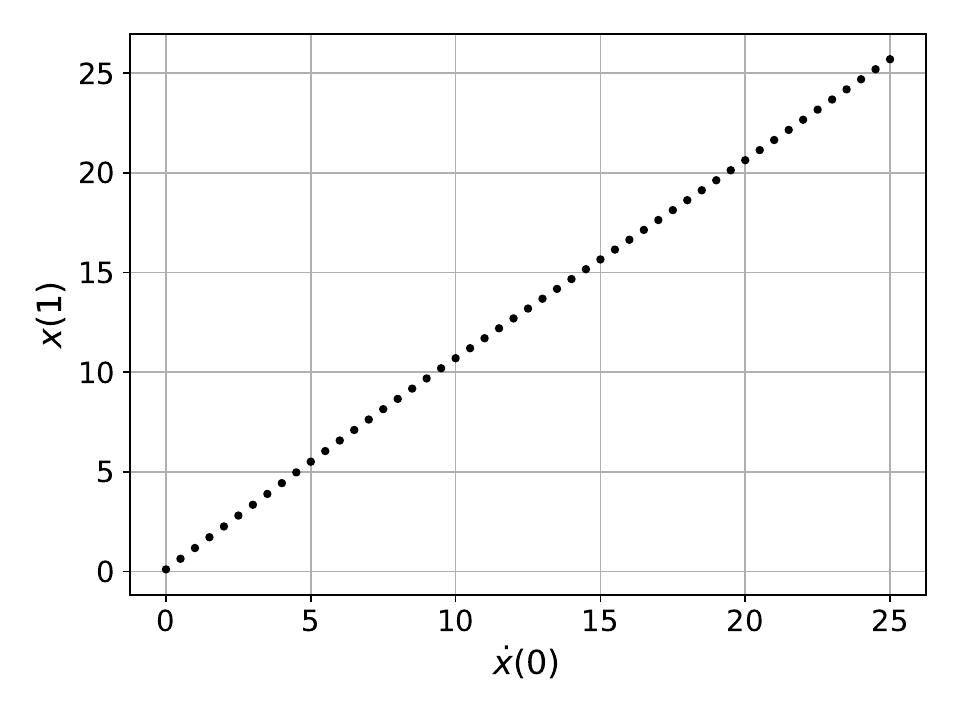}
  \caption{The prescribed friction force $F_{\text{True}}(x)$ used for data generation (left), and the corresponding final displacement $x(1)$ at $t = 1$, as a function of the initial velocity $\dot{x}(0)$ (right).}
  \label{fig:true_friction}
\end{figure}

Figure \ref{fig:true_friction} illustrates the non-trivial, spatially dependent friction force used in our study:
\begin{equation}\label{force}
  F_{\mathrm{True}}(x)= -\frac{1}{20000}(24 - x)(13 - x)^2(1 + x)^2 + 5 \,.
\end{equation}
The left panel displays the chosen form of $F(x)$, while the right panel shows the resulting trajectory at $t = 1$, $x(1)$, as a function of the initial velocity $\dot{x}(0)$.

\paragraph{Inverse problem for friction force and its analogy to holography.}
We now address the inverse problem, where the friction force $F(x)$ is the unknown quantity to be determined. The goal is to recover the presented force from measured data, such as the data shown in the right panel of Fig. \ref{fig:true_friction}. Specifically, this setup illustrates the conditions in \eqref{BCR} and \eqref{BCR2}:
\begin{align}\label{CMDL}
\begin{split}
\text{Boundary Condition:} \,\,\,\, x(0) =0   \,,\qquad 
\text{Data Loss Function:} \,\,\,\,  \{ \dot{x}(0) \,, x(1) \}  \,.
\end{split}
\end{align}

This problem directly mirrors the challenge in holography, where we attempt to recover hidden physical quantities (such as effective potentials) from boundary data. 

The analogy becomes particularly evident when considering the preparation of the data loss function used in holography. Recall that in the holographic context, we have data sets like $\{s, T\}$ for the QCD equation of state, and $\{T, \rho\}$ together with $\{T, s\}$ for holographic strange metals.

\begin{figure}
    \centering
    \includegraphics[width=0.45\linewidth]{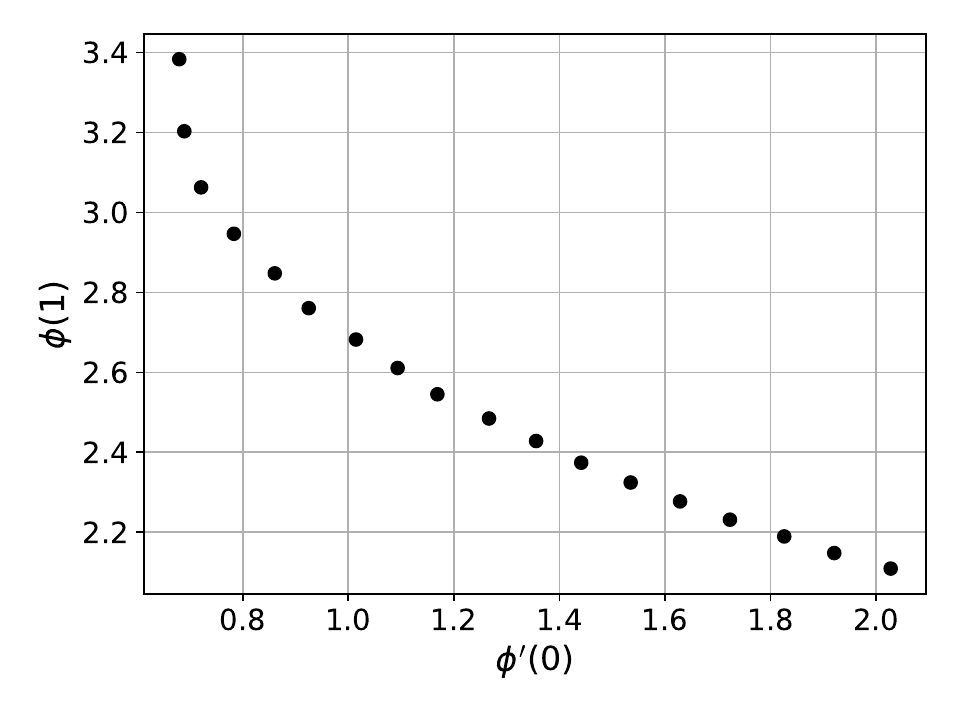}
    \includegraphics[width=0.45\linewidth]{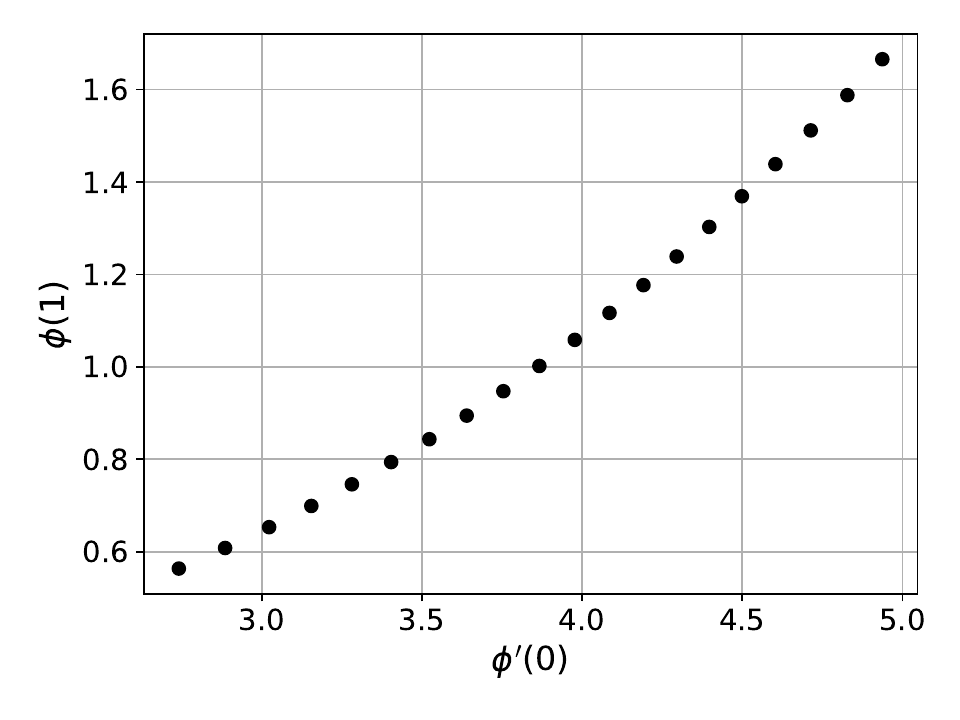}
    \caption{The training data $\{\phi'(0), \phi(1)\}$ for the QCD equation of state (left) and $T$-linear resistivity of holographic strange metals (right), which encode the same data of $\{T, s\}$ or $\{T, \rho\}$.}
    \label{fig:EoS_conceptual_data}
\end{figure}

To clarify this with the QCD example, we first solve the direct problem by computing the bulk equations of motion at a given potential $V(\phi)$, obtaining all bulk fields $f(z)$, $h(z)$, and $\phi(z)$. From these solutions, we can generate the data set for the inverse problem: $\{T, s\}$, which is equivalent to $\{f'(1), h(1)\}$. Alternatively, another combination of fields, such as $\phi(z)$, can also be used to generate the data loss function condition \eqref{BCR2}, such as $\{\phi'(0), \phi(1)\}$.\footnote{Heuristically, one can express this as $\phi(z=1;T) = \phi(z=1;s)$, where $s(T)$ is used to derive $s(\phi(z=1))$.} For instance, this relationship is illustrated in Fig. \ref{fig:EoS_conceptual_data}.\\

\begin{figure}[]
    \centering
    \includegraphics[width=0.32\linewidth]{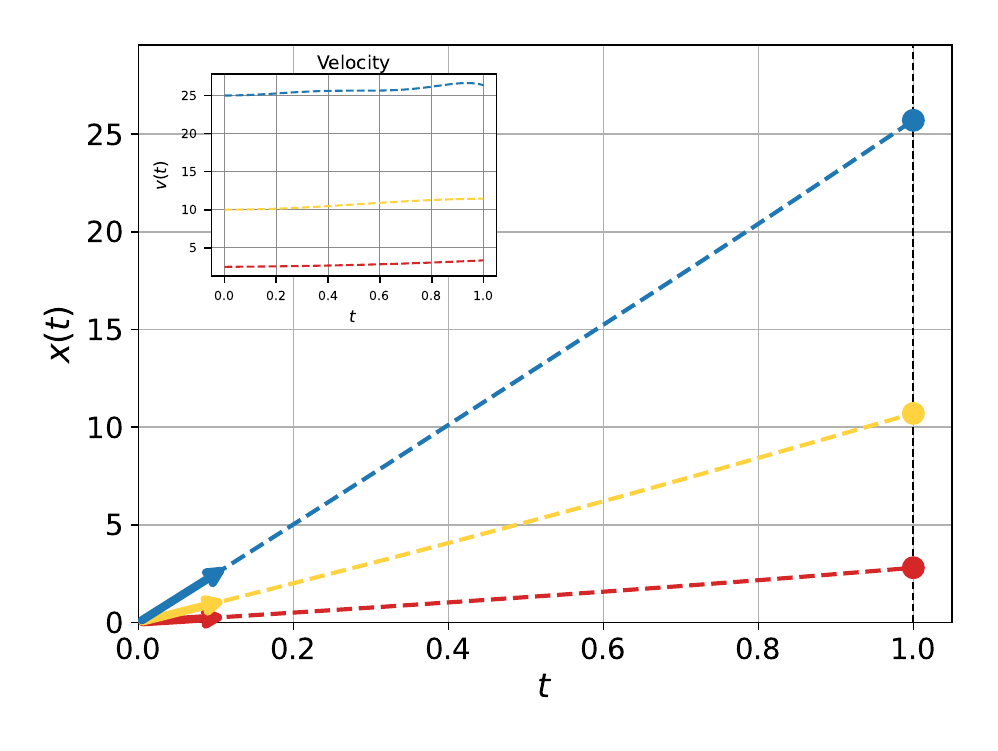}
    \includegraphics[width=0.32\linewidth]{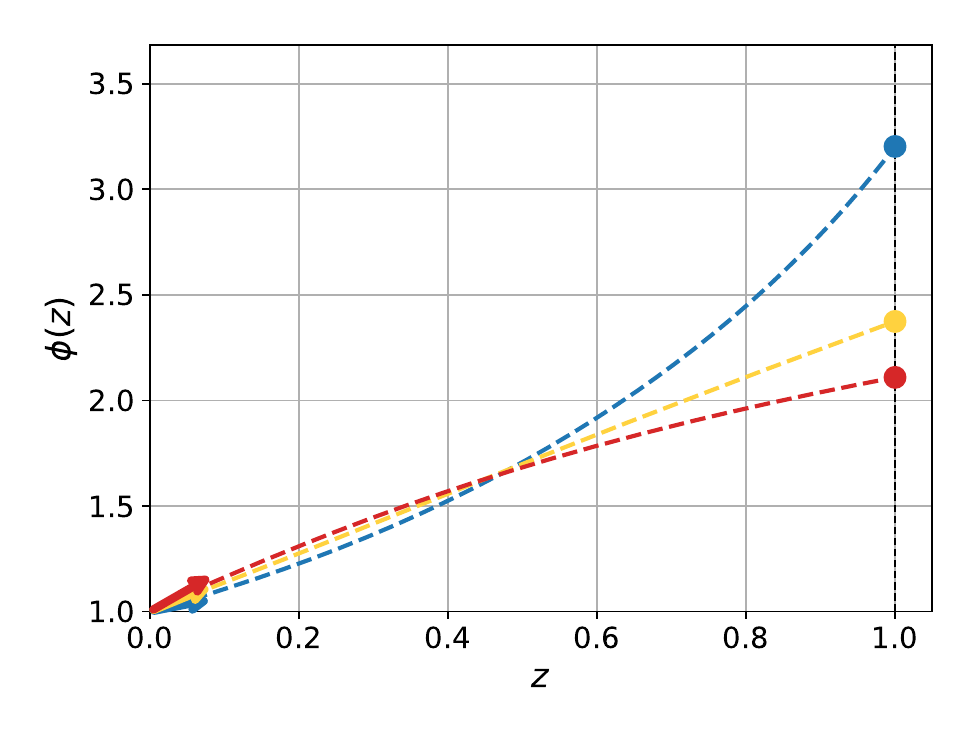}    
    \includegraphics[width=0.32\linewidth]{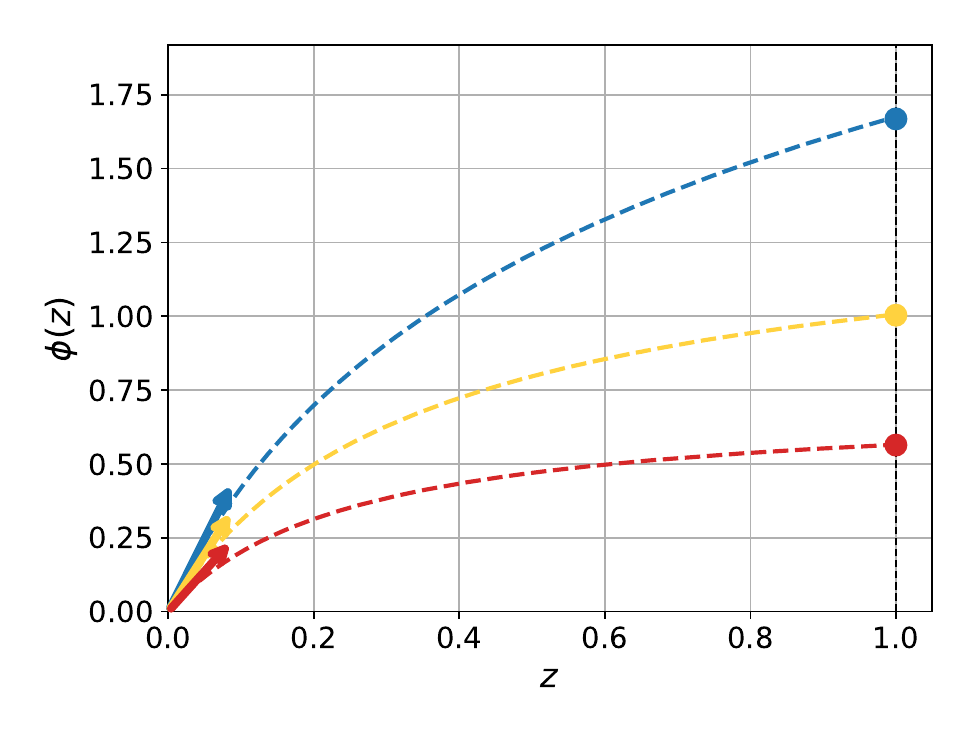}
    \caption{Trajectories of $x(t)$ and $\phi(z)$ depending on three different initial conditions $x'(0)$ or $\phi'(0)$, denoted by different colors: Classical mechanics (left), Holographic QCD (middle), and Holographic CMT (right).}
    \label{fig:Comparison_x_phi}
\end{figure}

The analogy can be summarized as
\begin{align}\label{}
\begin{split}
t \longleftrightarrow z \,,   \qquad
x \longleftrightarrow \phi \,, \qquad
F(x) \longleftrightarrow V(\phi) \,,
\end{split}
\end{align}
with the data loss function
\begin{align}\label{}
\begin{split}
 \{ \dot{x}(0) \,, x(1) \} \longleftrightarrow  \{ \phi'(0) \,, \phi(1) \} \,.
\end{split}
\end{align}
Figure \ref{fig:Comparison_x_phi} presents the direct analogy between the classical trajectory $x(t)$ and the holographic dilaton profile $\phi(z)$.

For the holographic CMT example, we consider two data sets: $\{T, \rho\}$ and $\{T, s\}$, which correspond to $T$-linear resistivity and specific heat in holographic strange metals. As in the QCD example, one can translate these input data sets into:
\begin{align}\label{}
\begin{split}
\{ \phi'(0) \,, \phi(1) \} \quad \text{and} \quad \{ \phi'(0) \,, \phi'(1) \} \,.
\end{split}
\end{align}
to determine the two unknown potentials $V(\phi)$ and $Z(\phi)$. The analogies between all variables and conditions are summarized in Table \ref{tab:Full_analogy} and Table \ref{tab:data_analogy}.
\begin{table}[]
\centering
{\renewcommand{\arraystretch}{1.5}
\begin{tabular}{c||c|c|c}
    & Classical mechanics & QCD EoS & $T$-linear resistivity \\
\hline\hline
Interval & $t \in [0,1]$ & $z \in [0,1]$ & $z \in [0,1]$ \\ 
\hline
Variables & $x$ & $\phi$, $f$, $h$ & $\phi$, $f$, $h$, $A_t$ \\
\hline
\begin{tabular}[c]{@{}c@{}} Constraints \end{tabular} 
& 
\begin{tabular}[c]{@{}c@{}} $x(0)=0$ \end{tabular} 
&
{\renewcommand{\arraystretch}{1.2}
\begin{tabular}[c]{@{}c@{}} $\phi(0)=1$, \\ $f(0)=h(0)=1$, \\ $f(1)=0$ \end{tabular}
}&
{\renewcommand{\arraystretch}{1.2}
\begin{tabular}[c]{@{}c@{}} $\phi(0)=0$, \\ $f(0)=h(0)=A_t(0)=1$, \\ $f(1)=A_t(1)=0$ \end{tabular}
} \\
\hline
Input data & $x'(0)$ & $\phi'(0)$ & $\phi'(0)$ \\
\hline
Output data & $x(1)$ or $x'(1)$ & $\phi(1)$ & $\phi(1)$ and $\phi'(1)$ \\
\hline
Unknown functions & $F(x)$ & $V(\phi)$ & $V(\phi)$ and $Z(\phi)$
\end{tabular}
}
\caption{Direct analogy between dynamical quantities in classical mechanics and holographic
inverse problems with QCD equation of state (EoS) and $T$-linear resistivity together with specific heat.}
\label{tab:Full_analogy}
\end{table}
\begin{table}[]
    \centering
    \begin{tabular}{c|c|c}
         &  Training Data  &  Target Function \\
    \hline
    \hline
      &  &  \\
      Classical mechanics  &  $\{x'(0),x(1)\}$ or $\{x'(0),x'(1)\}$  &  $F(x)$  \\
      &  &  \\
    \hline 
      &  & \\
      QCD EoS  &  $\{T,s\} \; \fallingdotseq \; \{\phi'(0),\phi(1)\} $  &  $V(\phi)$  \\
           &    & \\
    \hline
      &  &  \\
      $T$-linear resistivity  &  $\big\{\{T,s\},\{T,\rho\}\big\} \; \fallingdotseq \; \big\{\{\phi'(0),\phi(1)\},\{\phi'(0),\phi'(1)\}\big\}$  &  $V(\phi)$, $Z(\phi)$ \\
      &  &  \\
    \end{tabular}
    \caption{Analogy between training data and target functions in classical mechanics and holographic
inverse problems.}
    \label{tab:data_analogy}
\end{table}

\subsubsection{Data-driven identification of friction forces}
We now examine the inverse problem associated with the classical mechanics setup introduced earlier. To illustrate how PIML techniques can be used to identify unknown physical laws, we apply two deep‐learning approaches: Neural ODEs and PINNs.

To train these models, first we employ the data set $\{ \dot{x}(0) \,, x(1) \}$ shown in the right panel of Fig. \ref{fig:true_friction}, and we define the data loss used in Eq. \eqref{CMDL} as
\begin{equation}\label{}
    L_{\text{Data}} = \frac{1}{N}\sum_{i=0}^{N-1} | x_{\text{data}}^{(i)}(1) - x_{\text{trained}}^{(i)}(1) |  \,.
\end{equation}
where $N$ denotes the number of data point $\{ \dot{x}(0) \,, x(1) \}$ used in Fig. \ref{fig:true_friction}, which is $N=50$. Nevertheless, in principle, one may alternatively adopt a das loss function based on $\{ \dot{x}(0) \,, \dot{x}(1) \}$, the latter was found to be computationally more stable; all results below use this more robust formulation.

\paragraph{Neural ODE.}
Using the loss function above, we employ the Neural ODE framework of Sec. \ref{sec:2.1} to solve Eq. \eqref{EOM}, where the unknown friction law $F(x)$ is represented by a neural network $\mathcal{D}_F(x)$. The network consists of six hidden layers with 100 nodes each and ELU activations. For the numerical ODE solver, we use the Runge-Kutta method of order 5(4), DOPRI5~\cite{dormand1980family}, as implemented in the \texttt{PyTorch} library torchdiffeq package.

\begin{figure}[] 
    \centering
    \includegraphics[width=0.48\linewidth]{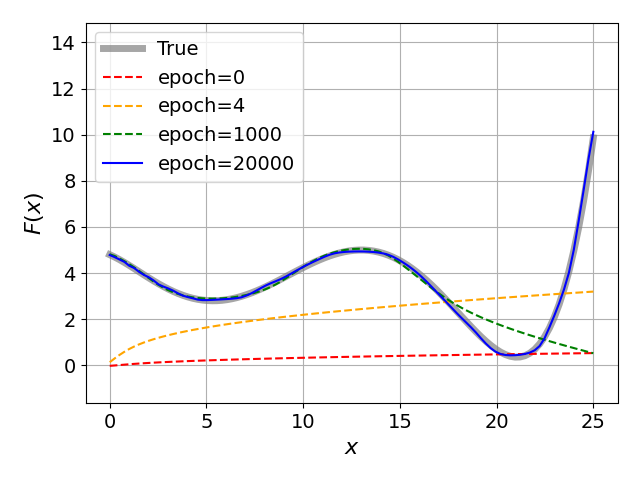}
    \hfill 
    \includegraphics[width=0.48\linewidth]{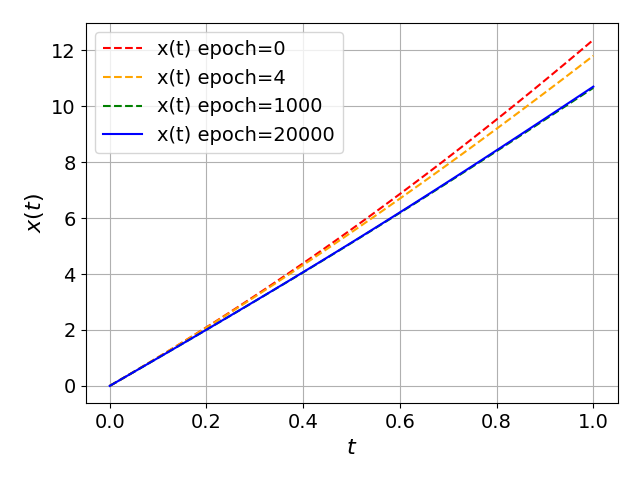}
 \caption{Neural ODE results for the friction force $F(x)$ and position $x(t)$ (with initial velocity $\dot{x}(0) = 10$) throughout the training epochs. The model's outputs steadily approach the true values as the training progresses.}
    \label{fig:neural_ode_comparison} 
\end{figure}

Training proceeds for 20,000 epochs, and the loss converges to $\mathcal{O}(10^{-2})$. Fig. \ref{fig:neural_ode_comparison} shows how the estimates of both the friction force $F(x)$ and the trajectory $x(t)$ improve throughout training. A clear sequential learning pattern emerges: once the model accurately reconstructs the position, it reliably identifies the friction law. This behavior persists for all tested initial velocities.

\paragraph{PINN.}
As an alternative model, we implement a PINN following Sec. \ref{sec:2.2}. Unlike Neural ODEs, PINNs incorporate both data loss \eqref{DATALOSSS} and physics loss \eqref{PINNLossEom} functions. The latter is given by the residual of the differential equation
\begin{equation}
\begin{split}
L_{\text{Physical}} = \frac{1}{N N_t} \sum_{i=0}^{N-1} \sum_{k=0}^{N_t-1} \left|  \ddot{x}^{(i)}(t_k) - g\sin\theta + \frac{F(x^{(i)}(t_k))}{m} \right|  \,.
\label{eq:loss_p}
\end{split}
\end{equation}
Both the friction force $F(x)$ and the trajectory $x(t)$ are parametrized by neural networks $\mathcal{D}_F(x)$ and $\mathcal{D}_x(t)$, with the same architecture as above (six layers, 100 nodes, ELU).

Several technical points arising in the PINN setting are in order. First, we impose the initial conditions by adopting
\begin{equation}
    x^{(i)}(t) = \dot{x}^{(i)}(0) \, t + t^2 \, \mathcal{D}_x^{(i)}(t) \,,
\end{equation}
which guarantees consistency with Eq. \eqref{CMDL} and also improves the convergence in the computations. Second, the physical loss function requires a discrete set of temporal points. We select $N_t=21$ uniformly spaced values in $t\in[0,1]$. Third, since the field $x(t)$ itself is a neural network, each trajectory $\mathcal{D}_x^{(i)}(t)$ implemented as a separate neuron in the output layer, following the discussion at the end of Sec. \ref{sec:2.2}.
\begin{figure}[]
    \centering
    \includegraphics[width=0.48\linewidth]{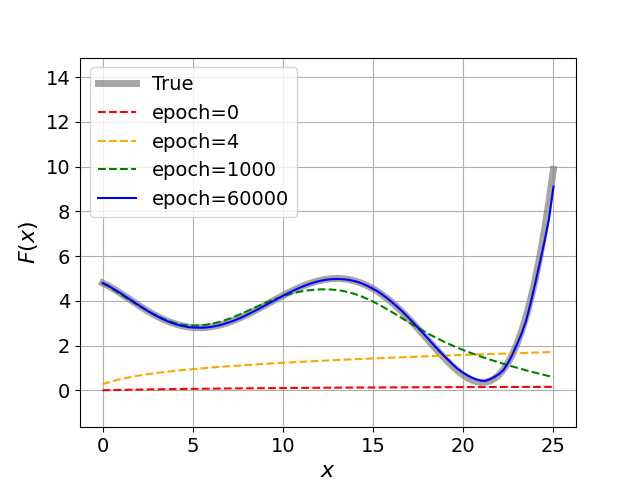}
    \hfill
    \includegraphics[width=0.48\linewidth]{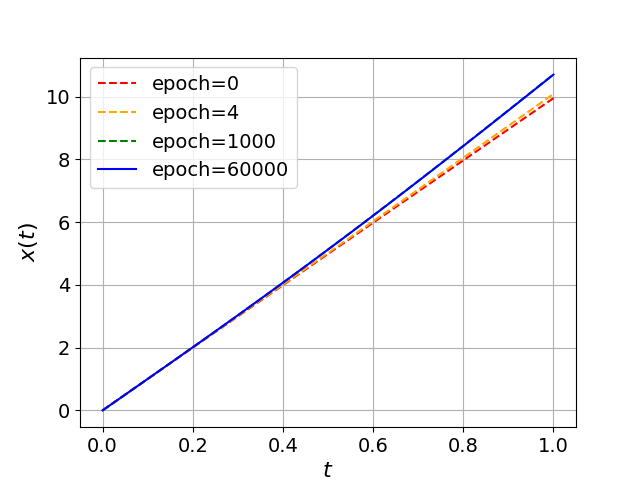}
 \caption{PINNs results for the friction force $F(x)$ and position $x(t)$ (with initial velocity $\dot{x}(0) = 10$) throughout the training epochs. The model's outputs steadily approach the true values as the training progresses.}
    \label{fig:PINN_comparison}
\end{figure}

The PINN is trained for 60,000 epochs and converges to a loss of order $\mathcal{O}(10^{-2})$. Fig. \ref{fig:PINN_comparison} depicts the progressive improvement of the learned friction force and trajectory, following the same sequential learning structure observed for Neural ODEs.\\

\begin{figure}[]
  \centering
  \includegraphics[width=0.45\textwidth]{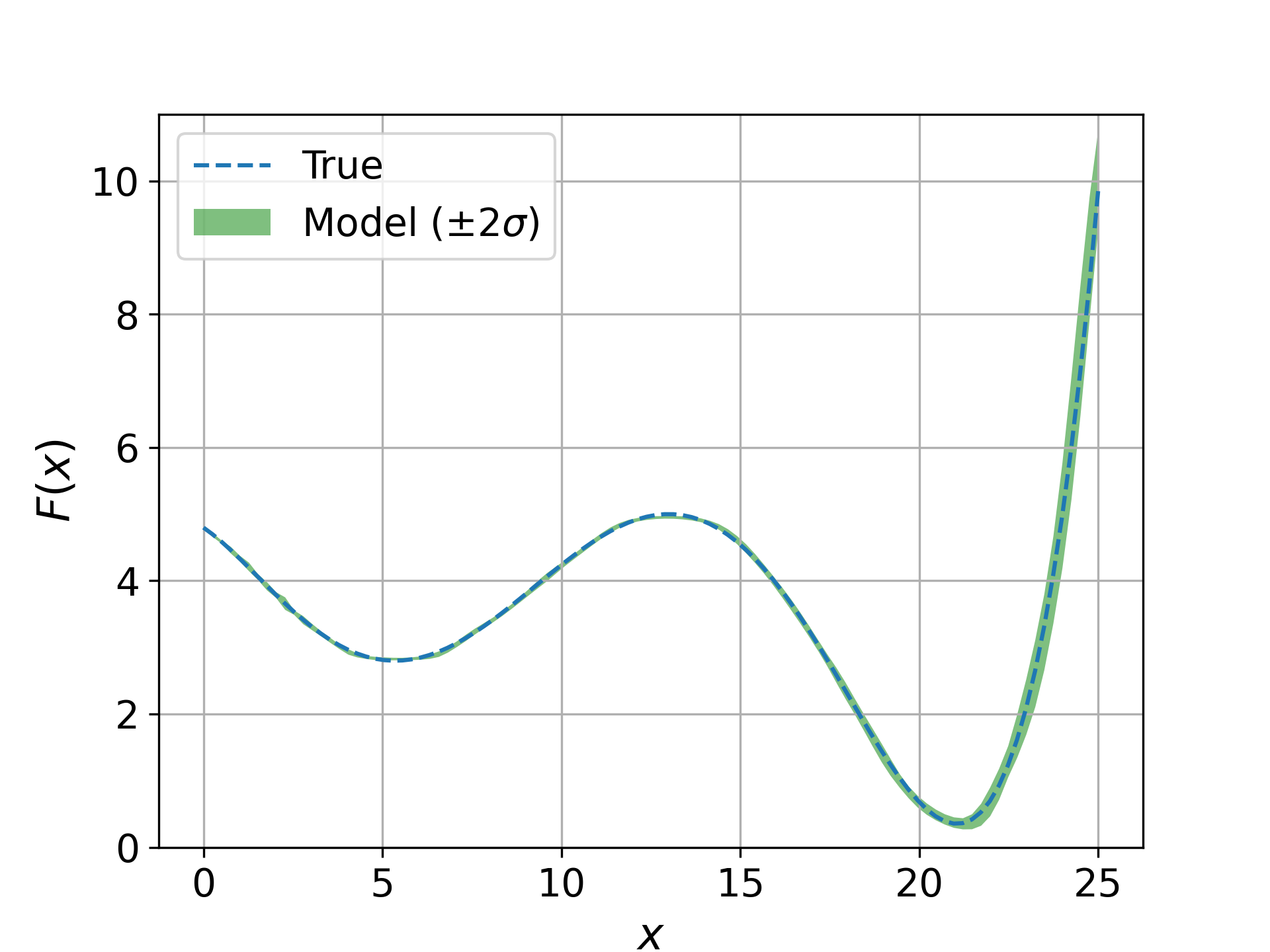}
  \includegraphics[width=0.45\textwidth]{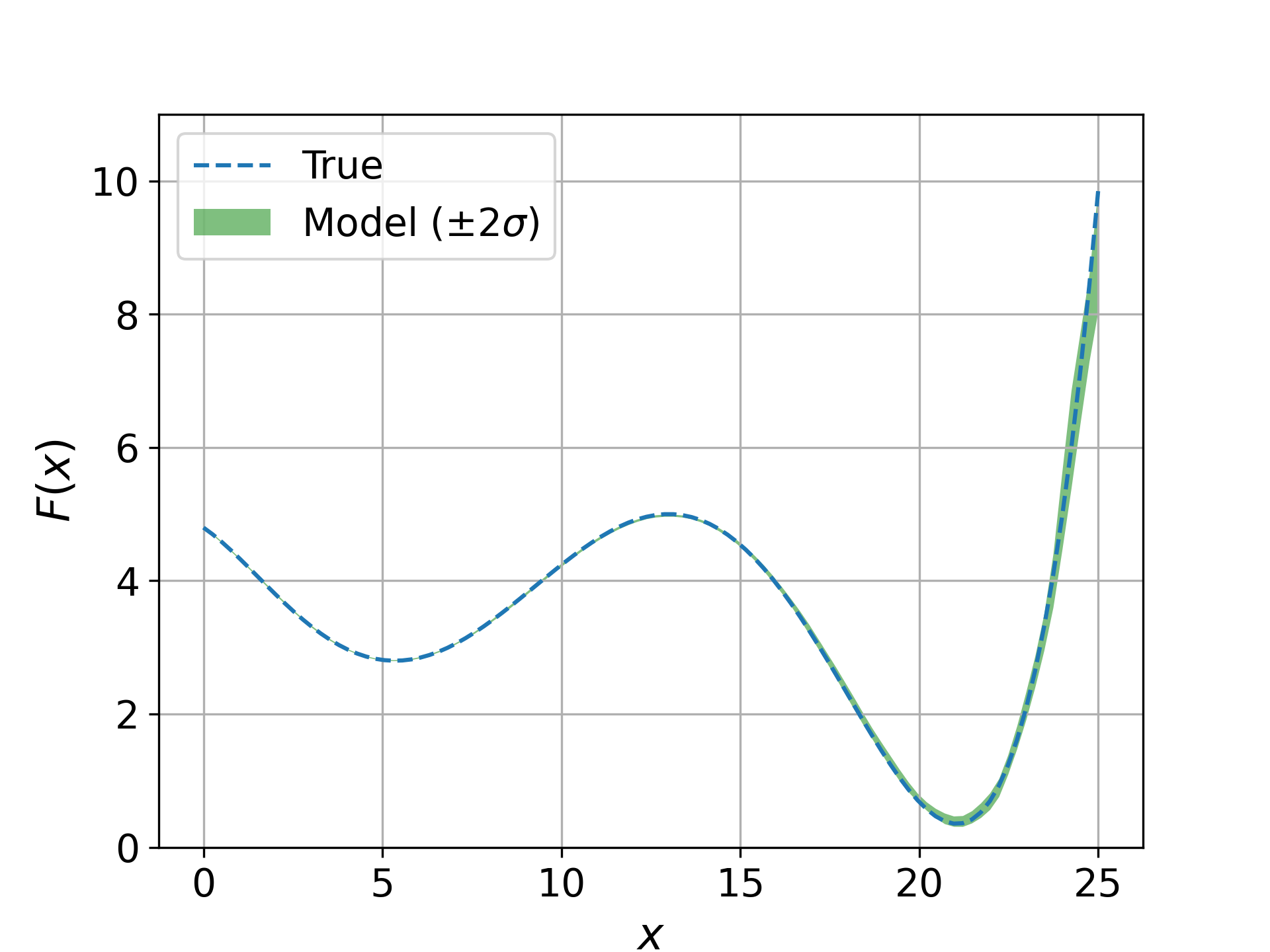}
  \caption{The inferred friction force $F(x)$ from the Neural ODE model (left) and the PINN model (right), where the dashed line represents the true friction force \eqref{force}. Both methods were trained across 20 trials, with the $2\sigma$ regions indicated.}
  \label{fig:model_predictions}
\end{figure}

Before concluding this section, we present the friction force reconstructed by both methods. Fig. \ref{fig:model_predictions} shows the inferred $F(x)$ after 20 independent training runs for each approach. Neural ODEs (left panel) and PINNs (right panel) both recover the true friction force with good accuracy, although PINNs exhibit slightly smaller variance $\sigma$ across runs. Furthermore, the total computation time for PINNs is empirically more than twice as fast as for the Neural ODE approach.
%

%
\subsubsection{Alternative to PINNs: Kolmogorov-Arnold Networks}

\begin{figure}
    \centering
    \includegraphics[width=1\linewidth]{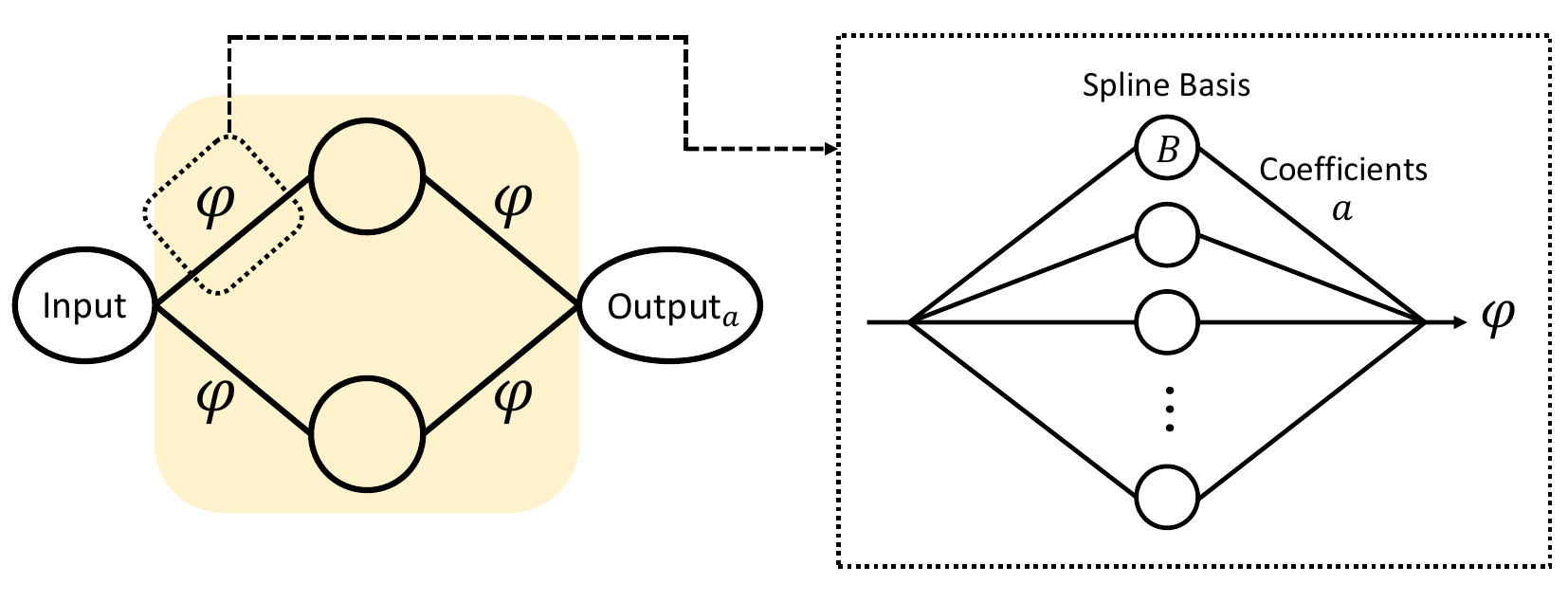}
    \caption{Schematic illustration of the KAN architecture. The diagram on the left shows a single-layer KAN with two hidden units, where the activation $\varphi$ is learnable on each connection.
    The diagram on the right shows that each activation 
    $\varphi$ is represented using a spline basis $B$ with trainable coefficients $a$.}
    \label{fig:kandiag}
\end{figure}

The Kolmogorov-Arnold networks (KANs) architecture~\cite{Liu:2024swq} is an alternative to the deep neural networks (DNNs) architecture. They are generally reported to achieve superior performance with the same layer and node configuration. In a given network structure, each edge is represented by a single-variable function parameterized by learnable parameters, and the outputs from preceding layers are combined at each node with a fixed weight of 1. For example, a KAN with a single hidden layer of 2 nodes, which is the architecture used throughout this paper, defines a function in the shape
\begin{equation}\mathcal{K}(\xi)=\sum_{j=1,2}\varphi^{(2)}_{1j}(\varphi^{(1)}_{j1}(\xi)) \,, \label{310}
\end{equation}
where the notation $\varphi^{(\ell)}_{ij}$ represents the function corresponding to the edge connecting the $i$-th node of the $\ell$-th layer and the $j$-th node of the $(\ell-1)$-th layer. The edge function is defined by 
\begin{equation}
    \varphi^{(\ell)}_{ij}(\xi)=\sum_{\alpha=1}^{n+d}a^{(\ell)}_{ij;\alpha} B_{\alpha,d}(\xi) \label{313} \;.
\end{equation}
This structure is visualized in Fig. \ref{fig:kandiag}.

In this model, each edge and corresponding single variable function itself is modeled by a linear combination of predetermined basis functions. In this paper, we use b-splines for the basis functions. On a fixed interval written $[0,1]$ without loss of generality, we start with defining the step functions (or zeroth degree functions)
\begin{equation}
    B_{i,0}(\xi)=\begin{cases}
        1:i/n \leq \xi \leq (i+1)/n\\
        0 : \text{otherwise}
    \end{cases}
\end{equation}
In order to obtain functions which $(d-1)$-th derivatives are continuous (or $d$-th degree functions) and are approximately zero outside the intervals $[i/n, (i+1)/n]$, we use the following recursive formula
\begin{equation}
    B_{i,d}(\xi) = \dfrac{\xi-i/n}{(i+p)/n-i/n}B_{i,d-1}(\xi)+\dfrac{(i+1)/n-\xi}{(i+1+p)/n-(i+1)/n}B_{i+1,d-1}(\xi).
\end{equation}
Note that due to each higher degree $B_{i,d}$ requiring the function $B_{i+1,d-1}$, naively each degree $d$ can define $n-d$ spline functions. To remedy this, one allows the values of $i$ for a $d$-th degree spline function to run from $-d$ to $n+d$, with $B_{i,0}=0$ for $i<0$ or $i>n$. This defines a total of $n+d$ spline functions at degree $d$. Then, each function correponding to an edge is associated with $n+d$ learnable parameters. For example, a single hidden layer KAN of 2 nodes with degree 3 and 5 grid points has a total of 8 parameters for each edge, and thus a total of 32 parameters.

\paragraph{Results of KANs.}
For the same physical scenario, we replace the DNN architecture with KANs in both frameworks, resulting in KAN Ordinary Differential Equations (KAN ODEs) and Physics-informed KAN (PIKAN) models. The unknown function $F(x)$ was approximated by $\mathcal{K}_F(x)$ with the single hidden layer of 2 nodes. Furthermore, we prepared another network $\mathcal{K}_x(t)$, which consisted of two hidden layers of 2 nodes and output layer of $N=51$ nodes corresponding the data index, for the PIKAN model. Thus, the solutions $x^{(i)}(t)$ are represented by $\mathcal{K}_x^{(i)}(t)$ which was $i$-th component of the vector function $\mathcal{K}_x(t)=(\mathcal{K}_x^{(0)}(t),\mathcal{K}_x^{(1)}(t),...,\mathcal{K}_x^{(N-1)}(t))$.

\begin{figure}[]
  \centering
  \includegraphics[width=0.44\textwidth]{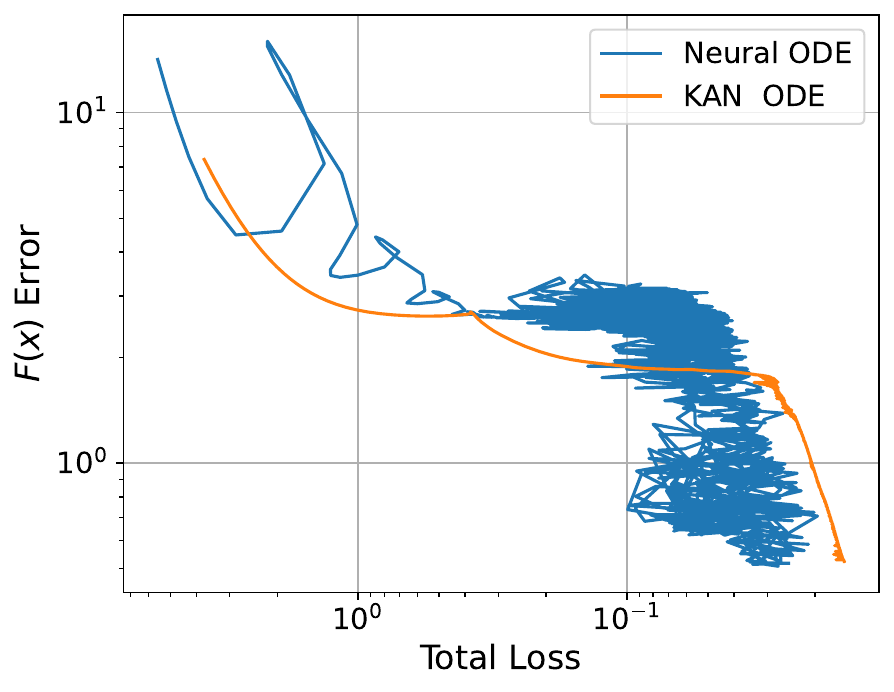}
\quad
  \includegraphics[width=0.45\textwidth]{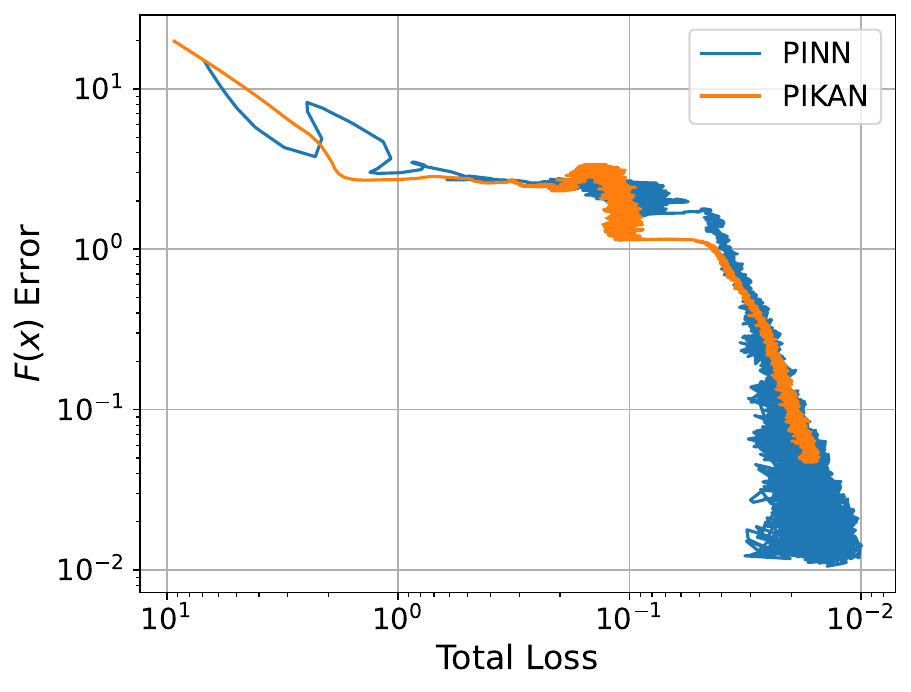}
  \caption{Mean squared error (MSE) of the reconstructed friction force $F(x)$ as a function of the total loss for different architectures. The left panel presents the results for neural ODE and KAN ODE. The right panel presents the results for PINN and PIKAN. In both cases, the KAN-based models show more stable training behavior with smaller loss fluctuations.}
  \label{fig:Err_F}
\end{figure}

We introduced the mean squared error (MSE) to examine the trend of agreement between the predicted function and the true function as following:
\begin{equation}
    \text{Error} = \frac{1}{25}\int_0^{25} | \mathcal{K}_F(x) - F_{\text{True}}(x) |^2 dx \;.
\end{equation}
As shown in Fig. \ref{fig:Err_F}, we compared the MSE of $F(x)$ with respect to the loss values between two architectures: DNN and KAN. This analysis provides insight into the stability of the training process. In both comparison settings, KAN ODE vs neural ODE and PIKAN vs PINN, the KAN-based models exhibited more stable training with smaller loss fluctuations, even at larger learning rates ($\mathcal{O}(10^{-2})$) than those of DNNs ($\mathcal{O}(10^{-3})$ and $\mathcal{O}(10^{-4})$). However, they required longer training time relative to its smaller network size.
%

%
\section{Conclusion}\label{sec4}
We have explored Physics-Informed Machine Learning (PIML), with a focus on Neural Ordinary Differential Equations (Neural ODEs) and Physics-Informed Neural Networks (PINNs) to address inverse problems in holography and classical mechanics. These neural network-based techniques are applied to reconstruct bulk spacetime and effective potentials in holographic models, as well as to model frictional forces in mechanical systems.

We demonstrate the successful use of these methods with two boundary quantum data examples: the QCD equation of state in holographic QCD and $T$-linear resistivity in holographic strange metals. Our results showcase the potential of PIML to solve complex, non-linear, data-driven physical systems traditionally challenging for analytical and numerical methods.

Additionally, we demonstrate that typical holographic problems can be directly analogous to classical mechanics, enabling the reformulation of gravitational dynamics as mechanical motion within an effective potential. This correspondence enhances both conceptual understanding and computational efficiency, expanding the applicability of PIML techniques beyond holography.

Our neural network-based methods, Neural ODEs and PINNs, enable the direct incorporation of physical data and laws into the loss function, facilitating the efficient reconstruction of hidden variables from boundary data: a key feature for interdisciplinary research. Additionally, we introduce Kolmogorov-Arnold Networks (KANs) as an alternative to traditional neural networks, demonstrating their superior performance in certain cases.

Beyond high-energy physics, the methodology developed here has broad implications for a variety of scientific and engineering domains, i.e., for AI-driven
scientific discoveries. The ability to embed physical laws directly into machine learning frameworks enhances model reliability and interpretability across fields such as mathematics, engineering, and the natural sciences.\\

While the results presented here demonstrate the power of PIML in solving inverse problems, there are several important directions for future research that could expand the applicability and efficiency of these techniques:

\paragraph{Improving model scalability and stability.}
While Neural ODEs and PINNs have shown effectiveness in solving inverse problems, their application to larger, more complex systems remains challenging. Future work should focus on improving the scalability of these models, particularly for systems with multiple interacting variables or highly irregular datasets. The extension of PIML techniques to quantum field theories and gravitational dynamics could lead to significant advances in understanding quantum gravity and addressing open problems in physics.

One promising direction is refining PINNs and exploring alternative architectures could lead to faster training and more accurate predictions. For instance, Bayesian physics-informed learning (B-PINNs), such as the B-XPINN, utilizes stochastic learning by replacing fixed network weights with Gaussian distributions~\cite{YANG2021109913,Hou:2024aa,LINKA2022115346}. B-PINNs have proven especially useful when dealing with limited or noisy data. Additionally, approaches like Extended PINNs (XPINNs) offer promising directions for further development in this area~\cite{D_Jagtap_2020,Hu:2021aa,HU2023107183,Dekhovich:2023aa}.

In addition, another promising avenue for future research is the integration of reinforcement learning with traditional PIML methods, creating hybrid models that can adaptively learn from both data and physical constraints. This could enhance the exploration of parameter spaces in complex systems and improve the robustness of the solutions.

\paragraph{Real-world holographic applications with PIML.}
As with many theoretical models, the ultimate test of these machine learning frameworks will be their application in experimental settings. Future studies should focus on validating these methods in laboratory contexts, especially in high-energy physics experiments, where real-world data could provide additional insights and further refine the models.

A promising area for further exploration is the application of PIML to strongly coupled systems, such as quark-gluon plasmas and other high-temperature phases. Investigating the dynamics of these systems using the frameworks discussed here could yield valuable insights into the behavior of matter under extreme conditions. Machine learning-driven models may also offer a more efficient approach to building more robust holographic systems, providing advantages over traditional numerical methods. In our ongoing projects~\cite{wipMLteam,wipMLteam2}, we are pursuing this direction by focusing on universal transport properties of these systems, such as the shear viscosity-to-entropy ratio of strongly interacting quark-gluon plasmas and Hall angle of non-Fermi liquids.\\

These directions represent promising avenues for future research, which we plan to address in the near future.

%
\acknowledgments
HSJ was supported by an appointment to the JRG Program at the APCTP through the Science and Technology Promotion Fund and Lottery Fund of the Korean Government. HSJ was also supported by the Korean Local Governments -- Gyeongsangbuk-do Province and Pohang City.
KYK was supported by the Basic Science Research Program through the National Research Foundation of Korea (NRF) funded by the Ministry of Science, ICT $\&$ Future Planning (NRF-2021R1A2C1006791), by the Ministry of Education (NRF-2020R1I1A2054376) and the AI-based GIST Research Scientist Project grant funded by the GIST in 2025.
All authors contributed equally to this paper and should be considered as co-first authors.

\providecommand{\href}[2]{#2}\begingroup\raggedright\endgroup

\end{document}